\documentclass[twocolumn,showpacs,aps,prd]{revtex4-1}
\usepackage{graphicx} 
\usepackage{float}
\usepackage{amsmath}
\usepackage{subfigure}
\usepackage{epstopdf}
\usepackage{multirow}
\usepackage{xspace}
\usepackage{rotating}
\usepackage{longtable}
\usepackage{multirow}
\usepackage[breaklinks=true]{hyperref}
\hypersetup{
  colorlinks=true,
  linkcolor=blue,
  citecolor=blue,
  urlcolor=blue
}

\graphicspath{{./fig/}}
\usepackage{array,tabularx,epsfig,mathrsfs,graphicx,rotating}
\usepackage{ifthen}
\usepackage{amsfonts}
\newcommand{\beq}{\begin{equation}}
\newcommand{\eeq}{\end{equation}}

\chardef\til=126

\newcommand{\zprime}{{\,\mathrm{Z^{0\prime}}}}

\newcommand{\met}{\mathrm{E_T\! \! \! \! \! \! \! / \; \; }}

\begin{document}

\title{\boldmath
Double Higgs production in the $4\tau$ channel from resonances in longitudinal vector boson scattering at a 100~TeV collider
}
\author{A.~V.~Kotwal}
\affiliation{
Fermi National Accelerator Laboratory, Batavia, IL 60510, USA  and Department of Physics, Duke University, Durham, NC 27708, USA 
}
\author{S.~Chekanov}
\affiliation{
HEP Division, Argonne National Laboratory,
9700 S.Cass Avenue,  
Argonne, IL 60439
USA
}
\author{M.~Low}
\affiliation{
Department of Physics,
University of Chicago,
Chicago, IL 60637,
USA
}

\hfill\draft{ }

\begin{abstract}
We discuss the sensitivity of a 100 TeV $pp$ collider to heavy resonances produced in longitudinal vector boson 
 scattering and  decaying to a pair of Higgs bosons. 
 A Monte Carlo study has been performed using the $H \to \tau \tau$ decay channel for both Higgs bosons, comparing the kinematics of such a signal to 
 the irreducible Standard Model backgrounds.     
The results are presented in the context of a phenomenological model of a resonance ($\eta$) coupling to goldstone modes, 
 $V_L V_L \to \eta \to HH$,  as can arise in composite Higgs models. 
With a fractional width of 70\% (20\%), the $5 \sigma$ discovery reach is  
 4.2 (2.9) TeV in resonance mass for 10 ab$^{-1}$ of integrated luminosity.  We also discuss the dependence of the mass reach on the collider energy and integrated luminosity.
\end{abstract}


\maketitle


\section{Introduction}

The discovery of the Higgs ($H$) boson~\cite{Aad:2012tfa, Chatrchyan:2012ufa} is one of the most exciting discoveries in physics. Its discovery 
 completes the particle spectrum predicted by the Standard Model (SM)~\cite{bib:Weinberg:1967tq,bib:Salam:1968rm}, and confirms the mechanism of
 spontaneous symmetry breaking to generate the longitudinal modes of the weak 
 gauge bosons~\cite{bib:Nambu:1960xd, bib:Goldstone:1961eq, bib:Goldstone:1962es, bib:Englert:1964et, bib:Higgs:1964ia, bib:Higgs:1964pj, bib:Guralnik:1964eu}. 
  However, the issue of very large radiative corrections to the mass of fundamental scalar particles has been acknowledged for some time~\cite{'tHooft:1979bh}. This is the famous
 ``naturalness'' or ``fine-tuning'' problem of the Higgs in the SM, where the quantum
 corrections due to quadratically diverging loop integrals tend to drive the parameter values towards a very high energy scale.
  Beyond-SM theories containing additional symmetries (such as supersymmetry or other global symmetries) 
 are able to protect the Higgs parameters from the quadratically-divergent radiative corrections. These symmetries imply the existence of new particles
 in loops which cancel or mitigate the loop corrections due to SM particles. 

 In a class of models~\cite{Georgi:1984af, Georgi:1984ef, Kaplan:1983fs}, 
 the SM Higgs boson  is itself another member of 
 a set of Goldstone modes, generated  from the spontaneous breaking of a larger global symmetry. Other Goldstone modes are the $W^\pm_L$ and the $Z_L$, which
 become the longitudinal
 components of the weak gauge bosons after electroweak symmetry breaking (EWSB). In a subset of these models~\cite{Dugan:1984hq, Agashe:2004rs, Contino:2006qr}, a common feature is 
 the spontaneous breaking of a global $SO(5) \to SO(4) $ symmetry, where the four Goldstone modes are the four real components of the $SU(2)_L$ doublet Higgs field.    

 The spontaneous breaking of the larger global symmetry is postulated to be caused by the formation of a condensate,
  due to  new strong dynamics at a high energy scale. As Goldstone bosons are automatically massless due to the Goldstone
 theorem, the question changes to why the Higgs boson mass is non-zero but small relative to the higher-energy
 compositeness scale. Various models have been proposed to explain this ``little hierarchy'' between the EWSB scale $v$ and  the new compositeness scale $f_\eta$. 

 A signature of this strong dynamics in the Higgs sector would be resonances coupling to the Goldstone modes, i.e. 
 the longitudinal $W$ and $Z$ bosons and the Higgs boson.  The lightest of such resonances would preferentially
  decay to the lightest particles interacting with this sector, ie. $W_L$, $Z_L$, and $H$. As a benchmark, we
 study the phenomenological model for a new scalar resonance $\eta$~\cite{Contino:2011np} 
 whose Lagrangian is
\begin{equation}
\mathcal{L} = \mathcal{L}_{\rm SM} + \frac{1}{2} \partial^\mu \eta \partial_\mu \eta - \frac{1}{2} m_\eta^2 \eta^2 + \frac{a_\eta}{f_\eta}\eta \partial^\mu \pi^a \partial_\mu \pi^a
\label{etaLagrangian}
\end{equation}
where $\pi^a$ represents the quartet of  Goldstone modes  in the Higgs doublet, $f_\eta$ plays the same role as the ``pion decay constant'' ($f_\pi$) in QCD chiral perturbation theory and
 $a_\eta$ is a dimensionless coupling. 

In composite Higgs models, rather than fully unitarizing longitudinal gauge boson scattering, the Higgs boson delays perturbative unitarity violation until the scale $f_\eta$.  At this scale the strong resonances appear which together unitarize the amplitude.  In the scalar resonance model we look at, there is one resonance $\eta$ below the compositeness scale which partially unitarizes the amplitude, depending on the coefficient $a_\eta$.  When $a_\eta = 1$ the $\eta$ completely unitarizes the amplitude, hence in this work we take $a_\eta = 1$ to simplify the high-energy behavior while
  characterizing  new strong dynamics in vector boson scattering (VBS)~\cite{Bagger:1993zf, Bagger:1995mk}. 
 Assuming no other decays other than to goldstones, in the high mass limit 
  the width of the $\eta$ resonance is given by~\cite{Contino:2011np}
\begin{equation}
\Gamma_\eta = \frac{a_\eta^2 m_\eta^3}{8 \pi f^2}
\label{etaWidth}
\end{equation}
implying that for our choice of $a_\eta = 1$ the fractional width is given by $\Gamma_\eta / m_\eta = m_\eta^2 / (8 \pi f^2)$. 

\begin{figure}[htp]
\centering
\includegraphics[scale=0.4]{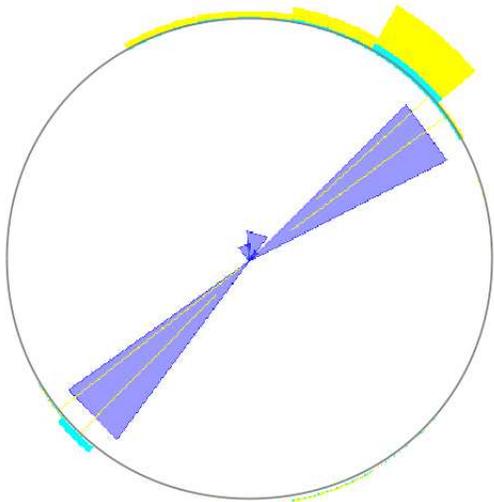}
\caption{A display of a simulated event generated using the process $pp \to \eta jj$ with the decay chain $\eta \rightarrow HH \rightarrow 4\tau$ at
a $100$ TeV $pp$ collider, with $m_\eta = 3$~TeV. 
Two Higgs-jets are shown with the large blue cones, each containing a pair of $\tau$ leptons.  Two small blue cones show the forward jets. The jets are 
reconstructed using the anti-$k_T$ algorithm \cite{Cacciari:2008gp}
 using the {\sc FastJet} package~\cite{fastjet}.
Yellow lines show charged hadrons.
}
\label{View1}
\end{figure}

 The sensitivity to the scattering process $W_L W_L \to \eta \to W_L W_L$ at a 100~TeV $pp$ collider  has been studied~\cite{Hook:2014rka, Cirelli:2014dsa, Bramante:2014tba, Curtin:2014jma,
 Craig:2014lda, Ismail:future, Berlin:2015aba}. The $W^+W^-$
 channel faces substantial background from $t \bar{t}$ production. It is interesting to probe $\eta$ production using both the
 $V_L V_L$ and $HH$ final states. The model predicts branching ratios in the proportions 2:1:1 for the $W_L W_L$, $Z_L Z_L$ and $HH$
 branching ratios based on counting the number of Goldstone modes in the Higgs doublet field.  The distinctive feature is that Goldstone modes
  have  purely derivative coupling, which preserves the shift symmetry of the Goldstone fields. Specifically, the coupling
 terms in Eqn.~\ref{etaLagrangian} expand to $\eta W^\mu_L W_{\mu L}$, $\eta Z^\mu_L Z_{\mu L}$, and $\eta \partial^\mu H \partial_\mu H$ respectively. 
 The complete interaction term between the $\eta$ resonance and the gauge bosons has the form $\eta [ 2g^2 W_\mu^+ W^{\mu -} + (g W_\mu^3 - g^{\prime} B_\mu)^2 ]$. 
 The specific Lorentz structure of these terms dictates the kinematic distributions associated with $\eta$ production and decay. 
  The combination of the information garnered from measuring the different branching ratios and the associated kinematic distributions 
  can provide the definitive test of the Goldstone nature of the Higgs doublet field and  information on its coupling to the new strong dynamics. Sensitivity to double-Higgs production in the
 context of the SM and other theoretical 
 approaches has been 
 investigated~\cite{Pierce:2006dh, Contino:2010mh, Contino:2012xk, Dolan:2012rv, Dolan:2012ac, Barr:2013tda, Dolan:2013rja, Chen:2014xra, Barr:2014sga, Azatov:2015oxa, No:2013wsa, Dawson:2015oha,Papaefstathiou:2012qe,Maierhofer:2013sha,deLima:2014dta,Goertz:2014qta,Papaefstathiou:2015iba,PhysRevD.91.073015,Li:2015yia}.  
The decay channels for the Higgs boson pair are indicated in Table~\ref{branchingRatios}, ranked by branching ratio~\cite{Agashe:2014kda}. In this table, we require that gauge bosons decay to leptons in order to 
 suppress 
 enormous backgrounds from QCD jet production. The first three channels in the table, $HH \to 4b$, $HH \to 2b2\tau$, and $HH \to \ell \nu \ell \nu bb$ are subject to large QCD backgrounds from 
  $b$-jets and  $t \bar{t}$ production 
 respectively. These considerations motivate the studies using the $4\tau$ and $2\gamma 2b$ final states. 
 The $2\gamma 2b$ final state has been studied both at 14 TeV~\cite{Baur:2003gp, Goertz:2013kp, Barger:2013jfa} and 100 TeV~\cite{Baglio:2012np, Yao:2013ika, Barr:2014sga, Azatov:2015oxa}
 due to its good mass resolution.   In this work, we present the first study of the $4\tau$ final state. 
 The $2 \gamma 2 b$ final state may be further studied in the future as a complementary channel with different backgrounds, providing additional discovery
 potential.

 \section{The $4 \tau$ Final State}

This analysis uses all decay channels for the $\tau$ leptons from the Higgs bosons. 
An example of the decay $\eta \to HH \to 4\tau$ is shown in Fig.~\ref{View1}.
It shows two Higgs bosons with transverse momenta above 1~TeV, arising from the decay of the $\eta$ boson with a mass of 3~TeV. The decay of each  Higgs boson leads
  to a single  jet containing two $\tau$ leptons.  
The event display was created using the {\sc Delphes} 
fast simulation \cite{deFavereau:2013fsa} and the Snowmass detector setup \cite{Anderson:2013kxz}.
The event, generated with the {\sc Madgraph5} 
Monte Carlo generator and showered with {\sc Pythia8} \cite{Sjostrand:2006za},
was taken from the {\sc HepSim} repository \cite{Chekanov:2014fga}.   A complete color version of this event is shown in the Appendix.

\vspace*{8mm}
\begin{figure}[htp]
\centering
\hspace*{6mm}
\includegraphics[scale=0.7]{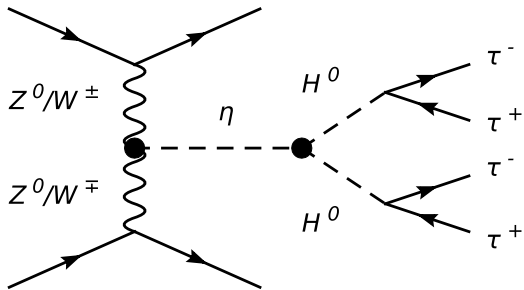}
\caption{Example of Feynman diagram for the process $pp \to \eta jj \to HH jj \to 4 \tau +jj$ via the production of the $\eta$ resonance in longitudinal vector boson fusion.  }
\label{etaDiagrams}
\end{figure}

This study is performed without simulation of the detector response, but we include the effect of the $\tau$-tagging efficiency  which is the most important
 performance characteristic. Hence it is likely that the inclusion of a realistic detector simulation with a similar $\tau$-tagging efficiency 
will not change significantly the results of this analysis. We assume that the requirement of four $\tau$ leptons will suppress mis-identification backgrounds to the level
 that SM processes producing four prompt $\tau$ leptons will dominate the backgrounds. This assumption is justified based on the $\tau$-lepton identification efficiency and QCD jet rejection
 achieved by the LHC experiments. For instance, the hadronic decays of the $\tau$-lepton are identified with an efficiency of 60\%, and with a QCD jet efficiency of 1-2\%, as reported 
 by the ATLAS experiment~\cite{ATLAS1, ATLAS2, ATLAS3, ATLAS4}. The analysis most similar to our analysis is the high-mass $\zprime \to \tau \tau$ search, where the transverse momenta ($p_T$)
 of the $\tau$ leptons are  similar to our signal kinematics. In the double-hadronic mode, 
 the dominant background arises from the $\gamma^*/Z \to \tau \tau$ Drell-Yan (DY) process, 
  followed by multijet and $W/Z$+jet background. The latter backgrounds are a factor of 3-4 smaller than the irreducible DY background. In the 
 leptonic+hadronic decay modes, the DY, $W$+jets, $t \bar{t}$, diboson and single-top backgrounds all contribute approximately equally. In order to estimate the fake backgrounds and compare them
 to the $ZZ$ background for our $HH$ search, we consider the diboson analysis for $ZV \to ll jj$~\cite{ATLAS5}. This analysis shows that requiring $Z \to ll$ and $V \to jj$ with a 
 mass window cut yields a $Z$+jets rate which is about 20-50 times larger than the $VV$ rate. 
 Given that the hadronic $\tau$ selection is about 15 times more efficient for prompt $\tau$'s compared to QCD jets, the
 requirement of two additional $\tau$'s will suppress the $Z$+jets background to a fraction of the $ZZ$ background. Multi-jet background will be reduced to a negligible level. Reference~\cite{ATLAS5} 
 also shows that the $t \bar{t}$ background is negligible in the high $p_T(V)$ region. The requirement of two additional $\tau$'s will suppress diboson+dijet and single-top+dijet
  backgrounds to a negligible
 level. 

 Dedicated studies with full simulation will be needed to design the future detectors which can maintain the $\tau$ identification performance at high $p_T(\tau)$ at the same level 
 that the LHC experiments have 
 demonstrated. The above discussion shows that, if this performance can be achieved, the sensitivity studies presented below using the irreducible backgrounds should provide a reliable
 estimate of the discovery potential.

\section{Monte Carlo simulations}

The goal of this   analysis is to estimate the discovery potential of a future 100 TeV scale 
 $pp$  collider, based on the expected event rates and distributions for the signal and backgrounds after kinematic and fiducial cuts.
The $\tau$-tagging efficiency, which is the main characteristic of detector performance,  is assumed to be 60\% similar to the LHC experiments~\cite{ATLAS1, ATLAS2, ATLAS3, ATLAS4}.

\begin{figure}[htp]
\centering
\includegraphics[scale=0.55]{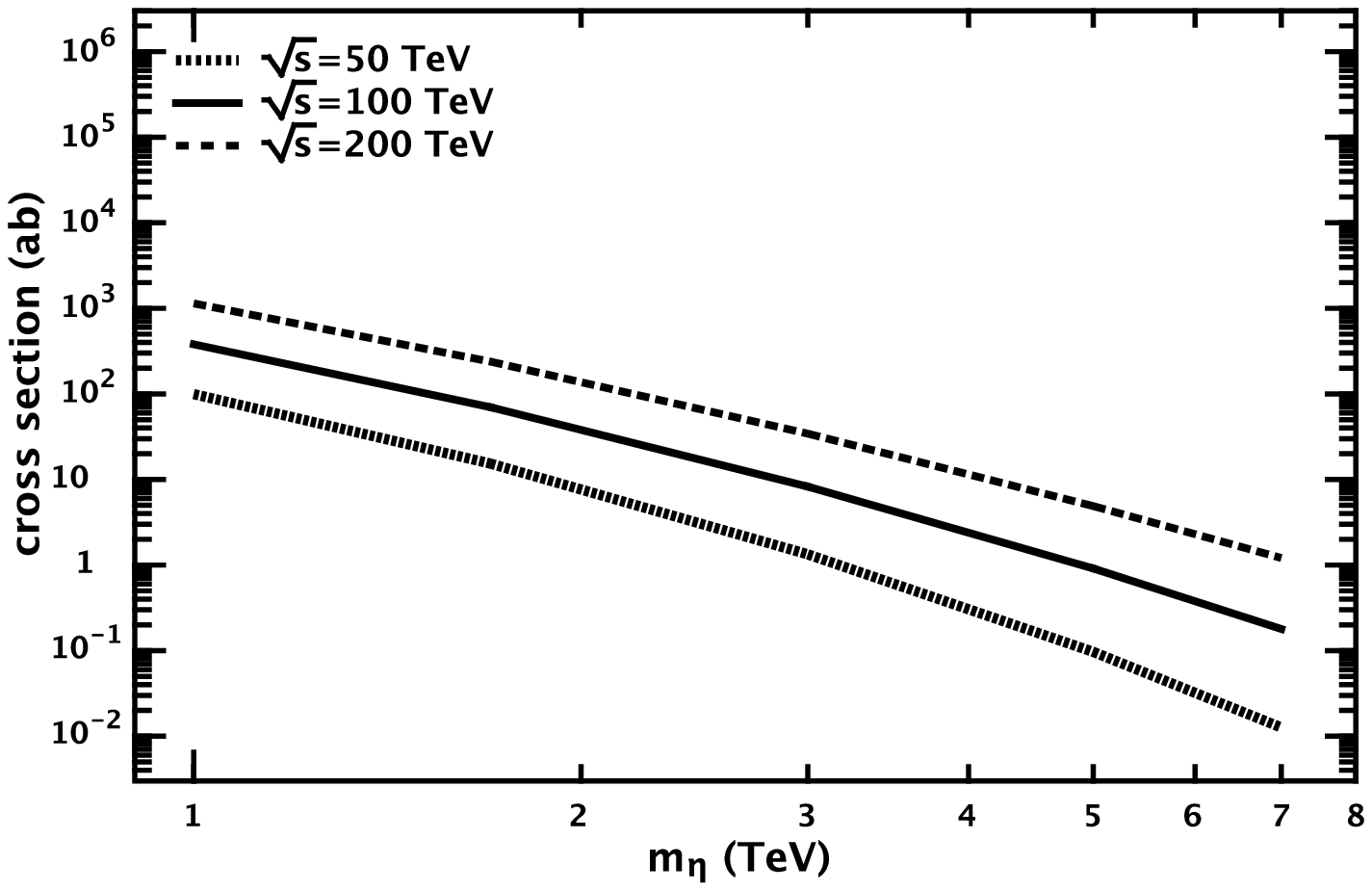}
\includegraphics[scale=0.55]{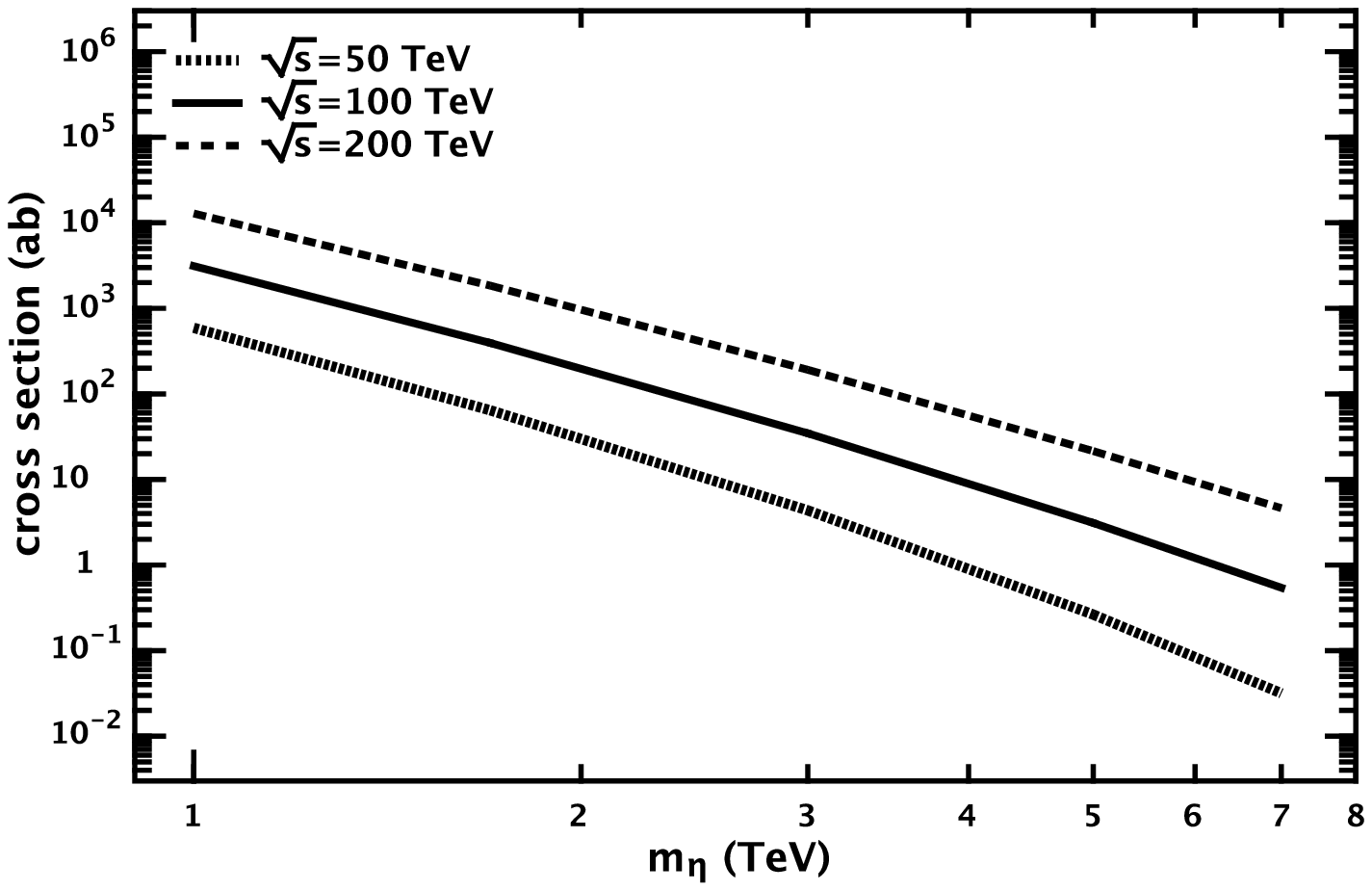}
\caption{Cross sections for the vector boson fusion  process $pp \to \eta jj$ with $\eta \to HH \to 4 \tau$
  for the fractional resonance width of 20\% (top) and 70\% (bottom), computed using the {\sc madgraph5} program at LO QCD. 
 The following generator-level cuts have been applied: $m_{jj} > 1.5$~TeV and $p_T (\rm jet) > 50$~GeV. 
}
\label{signalXS}
\end{figure}

The analysis was performed using the {\sc Pythia8}~\cite{Sjostrand:2006za} 
and {\sc Madgraph5}~\cite{Alwall:2011uj} 
MC models with the default parameter settings.
The MSTW2008lo68cl \cite{Martin:2009iq} parton density function (PDF) set was used.

The signal cross sections as functions of $m_\eta$ are shown in Fig.~\ref{signalXS}. These cross sections scale approximately as powers 
 of the resonance mass,  $m_\eta^{-a}$, where $3.2 < a < 4.8 $ ($3.9 < a < 5.2$) for a fractional resonance width of 20\% (70\%) and $\sqrt{s} = 100$~TeV.
  The range of $a$ shows the departure from a constant-power law
 as $1 < m_\eta < 7$~TeV, with the larger values of $a$ corresponding to the larger values of $m_\eta$. Thus, the value of $a$ can be used to estimate the slope of the curves in Fig.~\ref{signalXS}
  at different values of $m_\eta$.  The dependence of the cross section on collider energy may also be parameterized
 as a power law, $(\sqrt{s})^{b}$, with $1.8 < b < 3.3 $ ($2.2 < b < 3.6$) for a fractional resonance width of 20\% (70\%). Again, the range of $b$ corresponds to 
 $1 < m_\eta < 7$~TeV, with the larger values of $b$ corresponding to the larger values of $m_\eta$. 

\begin{figure}[htp]
\centering
  \subfigure[The pseudo-rapidity distributions of the forward jets.]{
  \includegraphics[scale=0.45, angle=0]{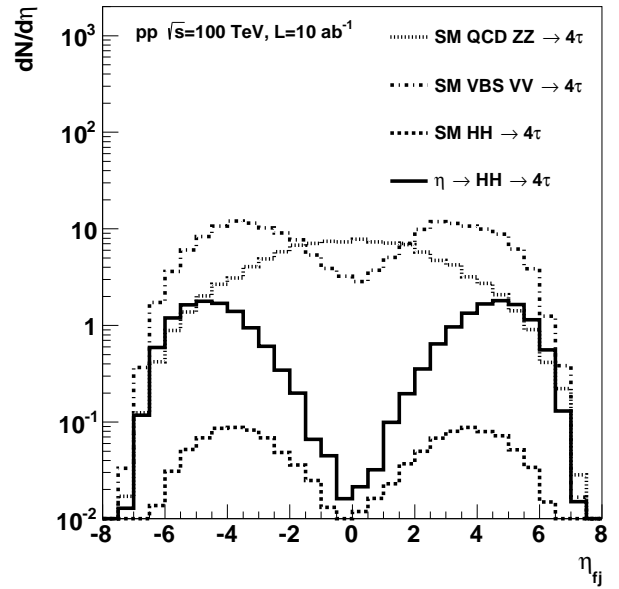}
  }
  \subfigure[The pseudo-rapidity distribution of the forward jet with larger absolute pseudo-rapidity.]{
  \includegraphics[scale=0.45, angle=0]{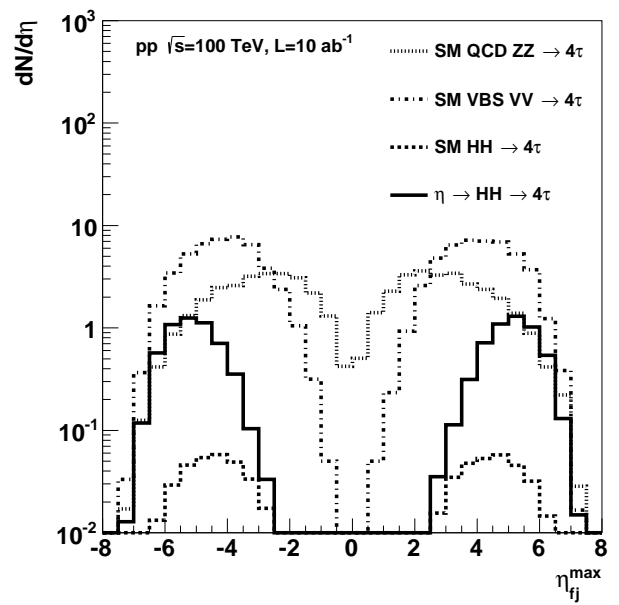}
  }
\caption{The pseudo-rapidity distributions of the two forward jets. Generator-level cuts $m_{jj} > 1.5$~TeV and $p_T({\rm jet}) > 50$~GeV have been applied on the jets for the samples shown. 
 Furthermore, the generator-level cuts on the samples in all figures include 
  $p_T(\tau)> 100$~GeV,  $p_T^{\rm leading}(\tau) > 300$~GeV and $| \eta (\tau) | < 3$. 
 The signal distribution in all figures corresponds to $m_\eta = 3$~TeV with a fractional width of 20\%. 
}
\label{fj_eta}
\end{figure}

 The {\sc HepSim} public repository \cite{Chekanov:2014fga} was used to  store  simulated events in the {\sc ProMC} file format
\cite{2013arXiv1306.6675C,Chekanov20142629}. The  samples
were analyzed  with a C++/ROOT program \cite{root}.
The jets were reconstructed with the anti-$k_T$ algorithm \cite{Cacciari:2008gp}
 using the {\sc FastJet} package~\cite{fastjet}. The typical choice of the distance parameter $R$ for jet reconstruction is $R \sim 0.4$ for light-quark and gluon jets, 
 and this value is motivated for the reconstruction of the forward jets. A smaller value of $R \sim 0.05$ is motivated for the reconstruction of the highly-boosted $\tau$-jets present in our
 samples. For simplicity, a single value of $R = 0.2$ is used in this study, since the kinematic distributions are fairly broad and the discrimination between signal and background is
  not sensitive to optimization of the choice of $R$. 
  We assume that the development of sophisticated $\tau$-reconstruction algorithms and the use of
 jet sub-structure information in the ultimate data analysis will permit the separation of boosted $H \to \tau \tau$ and $Z \to \tau \tau$ jets from mis-identified QCD jets and 
 electrons mimicking $\tau$-jets. 

 Jets have a generator-level requirement of $p_T (\mathrm{jet})>50$~GeV, based on studies performed for the high-luminosity (HL-) LHC~\cite{ATLAS-HLLHC-VBS, CMS-HLLHC-VBS} where this requirement was applied
  to suppress pileup jets
 in the forward region to an acceptable level. Given the hard dijet-mass ($m_{jj}$) spectrum arising from the VBS topology, a generator-level requirement of $m_{jj} > 1.5$~TeV 
 has been applied to increase the event generation efficiency. 
 For jet clustering, stable particles with lifetimes greater than $3\cdot 10^{-11}$ seconds are selected, and neutrinos are excluded.

\begin{figure}[htp]
\centering
  \subfigure[The distribution of the difference  $\Delta \eta_{jj}$ in pseudo-rapidities  of the two forward jets.]{
\includegraphics[scale=0.45, angle=0]{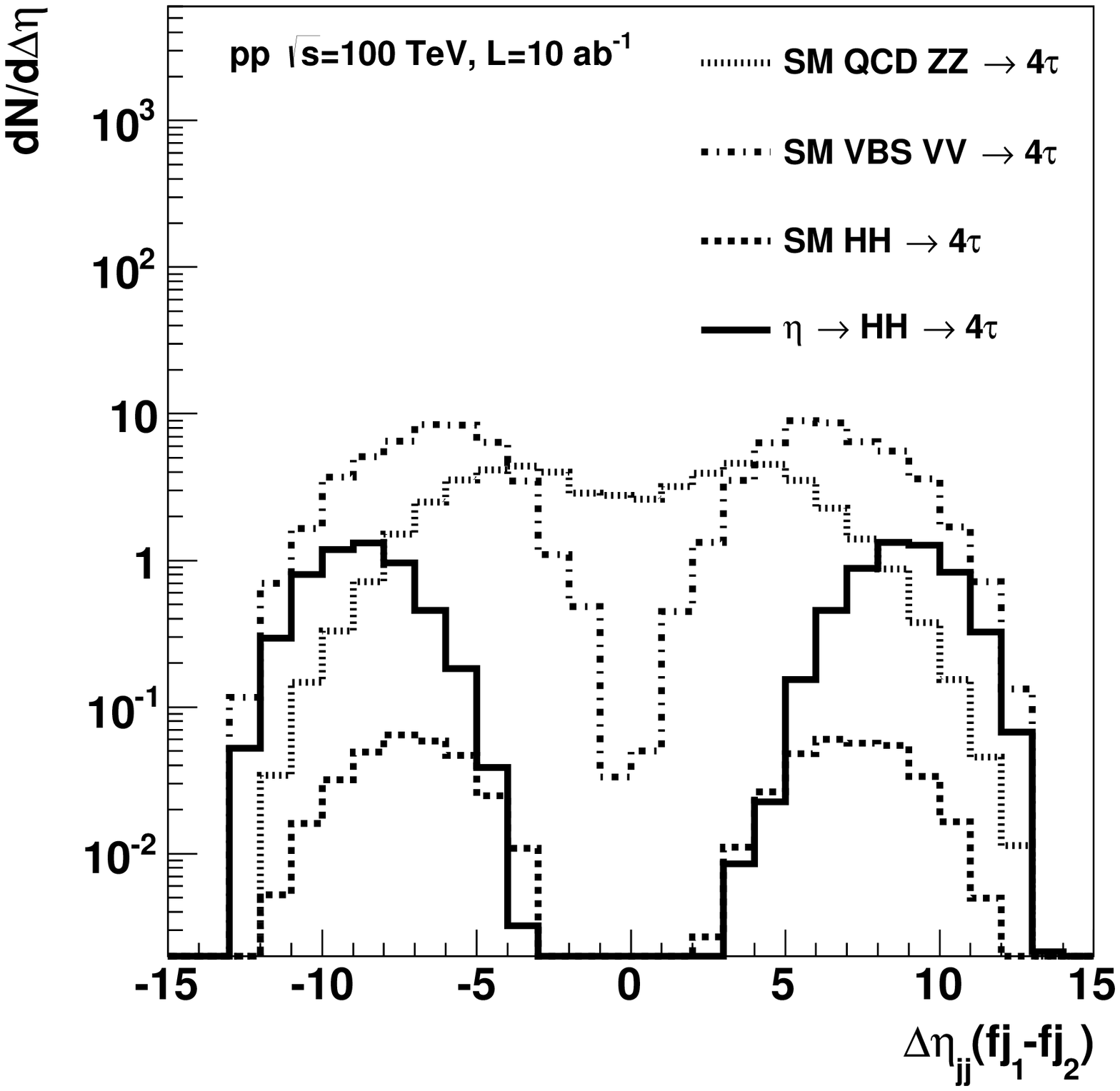}
}
  \subfigure[The distribution of $H_T^c$, the scalar sum of the $p_T$ of the central jets. ]{
\includegraphics[scale=0.45, angle=0]{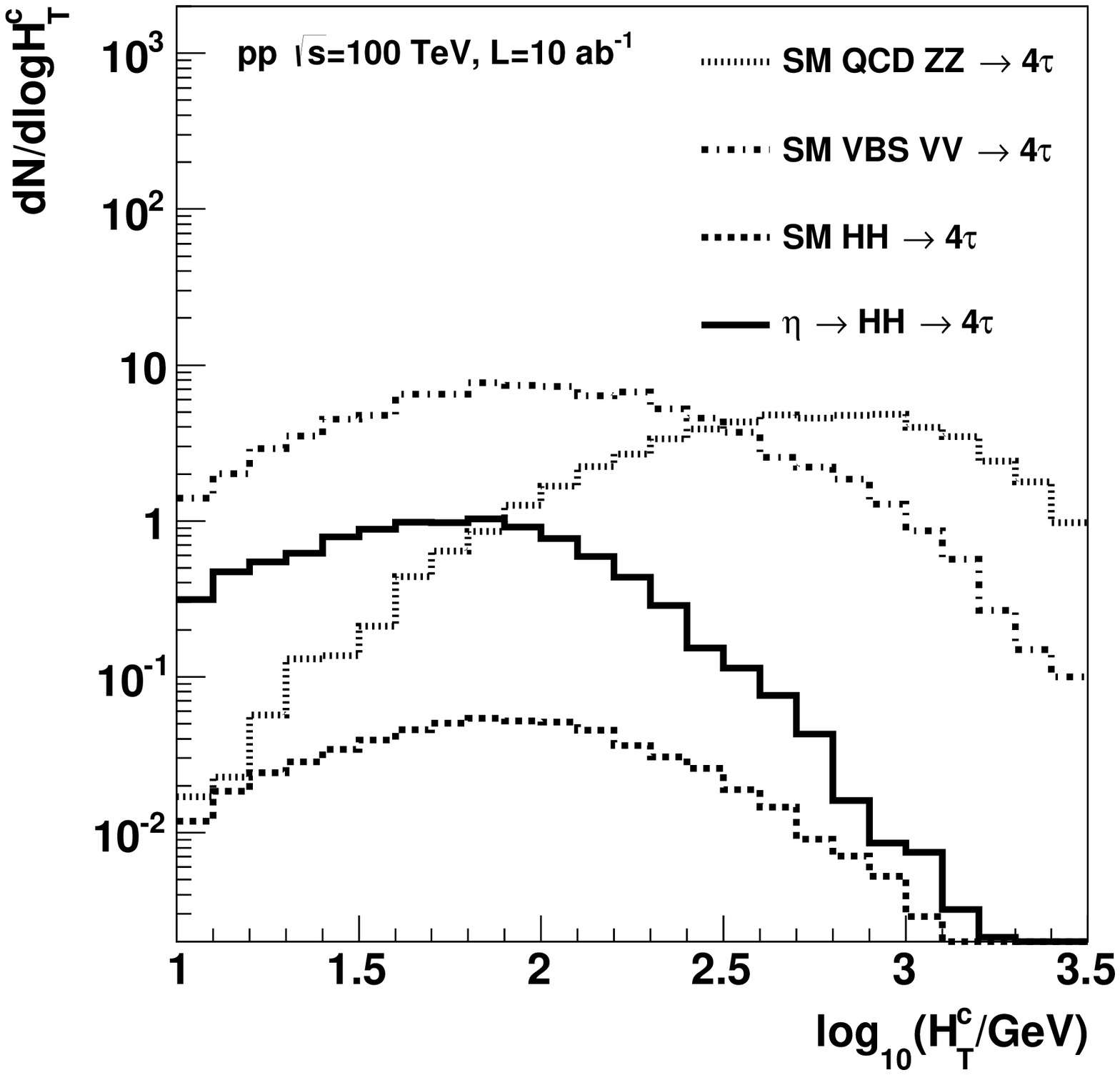}
}
\caption{The samples shown in this figure  include the generator-level cuts $m_{jj} > 1.5$~TeV, and 
 $p_T({\rm jet}) > 50$~GeV. The distributions motivate the additional requirements of $| \Delta \eta_{jj} | > 5$ and  $H_T^c < 300$~GeV.}
\label{fj_deta}
\end{figure}

The SM background  predictions were performed at leading order (LO) in QCD with the {\sc Madgraph5} program.  
 We include the following irreducible background processes in this study; (i) $VVjj \to 4\tau jj$ production ($V = Z, \gamma^*$) via purely electroweak couplings,  (ii) $ZZ jj \to 4 \tau jj$ production 
via the presence of the strong coupling in the Feynman amplitudes,  
 and (iii) $HH jj \to 4 \tau jj$ production via purely electroweak couplings. 
 The $VV$ production from the vector boson scattering topology  was computed separately from 
  $ZZ$ events where the jets are radiated from QCD vertices.   
 The interference between the $ZZ jj$ amplitudes with and without QCD vertices has been shown to be about 7\% in the relevant phase space
 at the LHC~\cite{ATLAS:2014am}. We consider this interference effect negligible for the purposes of this study. We have also neglected $HHjj$ production via gluon fusion as this contribution
 is suppressed by the selection requirements favoring the VBS topology. 
 The transverse momenta of the generated $\tau$ leptons were required to be $ p_T(\tau) > 100$~GeV, with the leading $\tau$ lepton required to have $p_T (\tau) > 300$~GeV, and all  $\tau$ leptons 
 are required to have
  pseudo-rapidity $|\eta(\tau)|<3$,  to increase the efficiency for event generation. 

 Next-to-leading order QCD $k$-factors have been estimated~\cite{vbfnlo,marcAndre} to be  $\sim 50$\% for QCD production of $VVjj$ and $<10$\% for VBS.  As the latter 
 background is dominant in this analysis, the relevant NLO-QCD correction is both small and similar for the signal VBS topology. Thus, NLO-QCD corrections are expected to have a 
 negligible impact on the results. 

 A potential additional source of signal is the $ZZ$ decay mode of the $\eta$ resonance, via the $\eta \to ZZ \to 4\tau$ channel. Due to the factor of $\sim 2$ 
 smaller branching ratio for $Z \to \tau \tau$
 compared to $H \to \tau \tau$, this contribution is small compared to the signal we have considered. For simplicity we have neglected this contribution, yielding conservative results for 
 signal sensitivity. Alternatively, the $\eta \to ZZ \to 4\tau$ channel may be distinguished from the $\eta \to HH \to 4\tau$ channel using advanced analysis techniques, and combined with 
 the $Z \to ee / \mu \mu$ channels to check the branching ratios of the resonance to the Goldstones. 
 
\begin{figure}[htp]
\centering
  \subfigure[The $p_T$ distribution of forward jets. ]{
  \includegraphics[scale=0.45, angle=0]{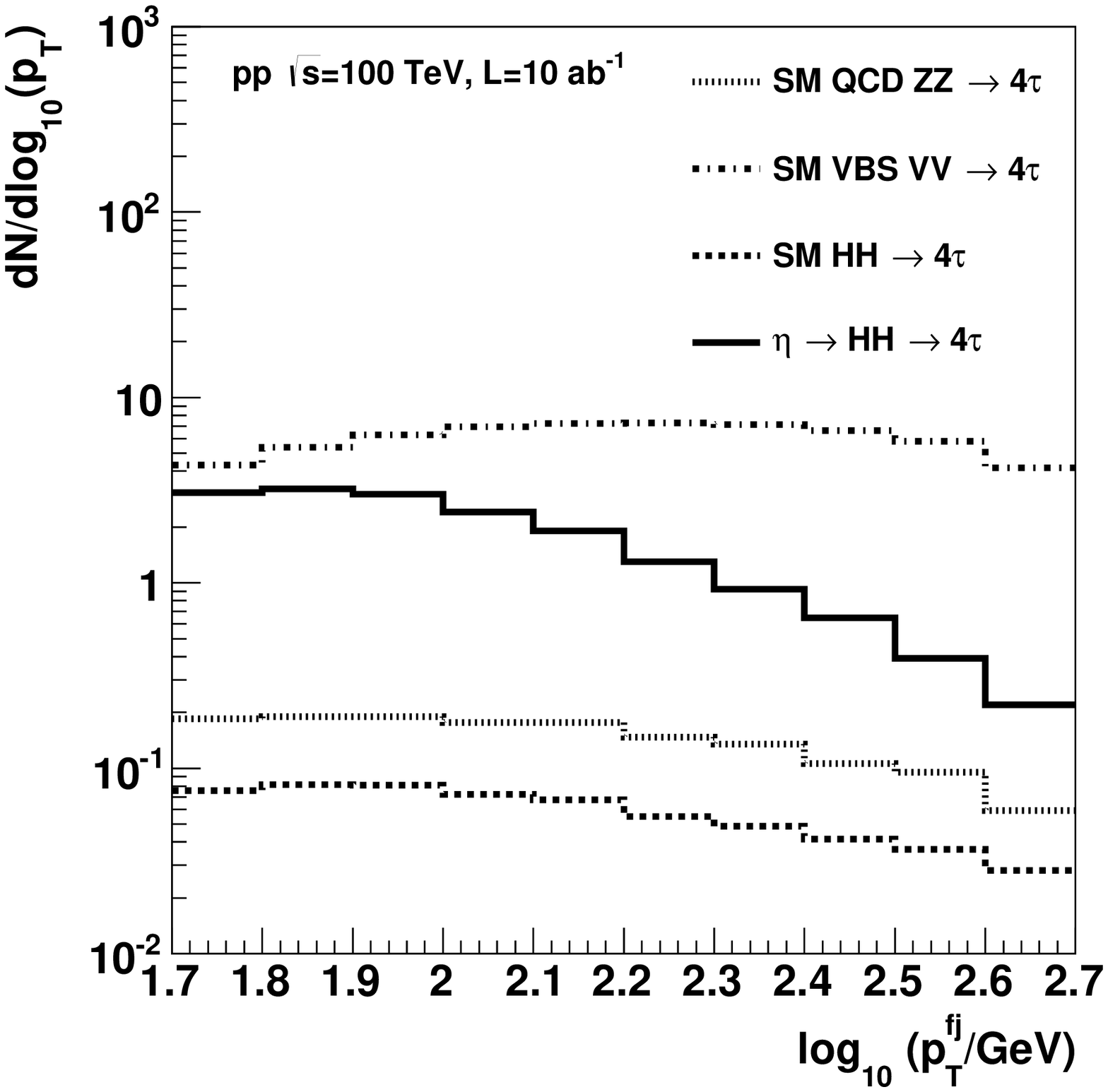}
  }
  \subfigure[The distribution of dijet mass of the  forward jets. ]{
  \includegraphics[scale=0.45, angle=0]{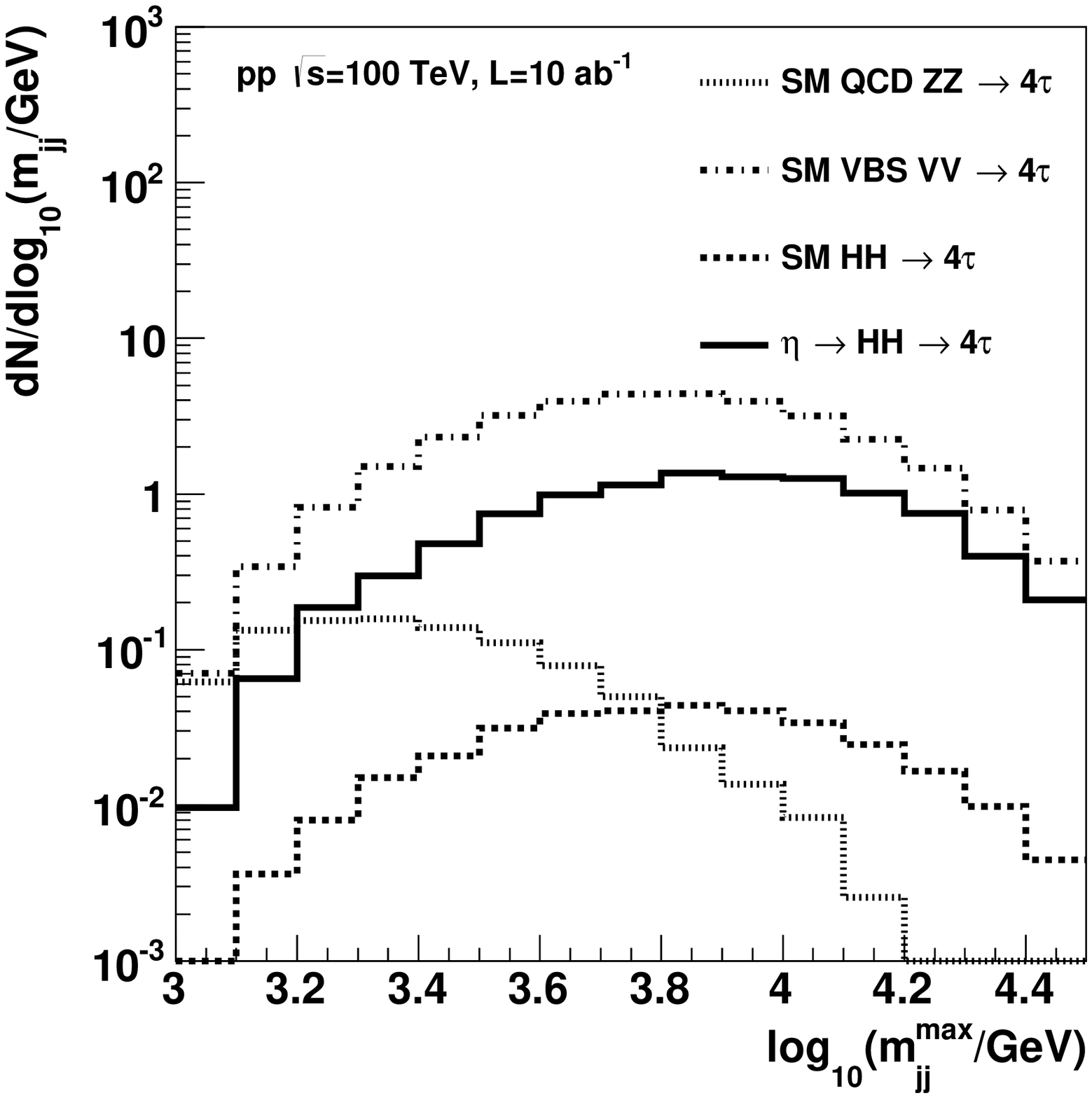}
  }
\caption{The $p_T$ and dijet mass distributions of the two forward jets.  The samples shown in this figure and subsequent figures include the generator-level cuts $m_{jj} > 1.5$~TeV,
 $50 < p_T({\rm jet}) < 500$~GeV, $| \eta({\rm jet}) | > 2$, $| \Delta \eta_{jj} | > 5$ and $H_T^c < 300$~GeV.
}
\label{fj_pt}
\end{figure}

\section{Kinematic Distributions}

 Figure~\ref{fj_eta} shows the pseudo-rapidity distribution of the forward jets and the distribution of pseudo-rapidity of the forward jet with larger absolute pseudo-rapidity.  
  Forward jets are defined as the jet pair with the  largest invariant mass $m_{jj}$ in the event. Studies at the LHC have shown
 this criterion to be effective in identifying the forward tagging jets in VBS. The pseudo-rapidity of the more-forward jet gives an indication of the required rapidity coverage of the detector. 
 The figures show that processes involving QCD vertices produce jets which are more central, while the VBS topology for SM and $\eta$ resonance processes tends to produce jets in the forward
 direction. Furthermore, the $\eta$ resonance process is mediated by longitudinal VBS, where the scattered quarks emerge at higher rapidities as compared to the SM VBS process mediated predominantly 
 by transverse vector bosons. Thus, new physics that is specific to longitudinal vector bosons will be a primary driver for maximizing forward rapidity coverage. 

\begin{figure}[htp]
\centering
  \subfigure[The pseudo-rapidity distribution of $\tau$-jets.]{
  \includegraphics[scale=0.45, angle=0]{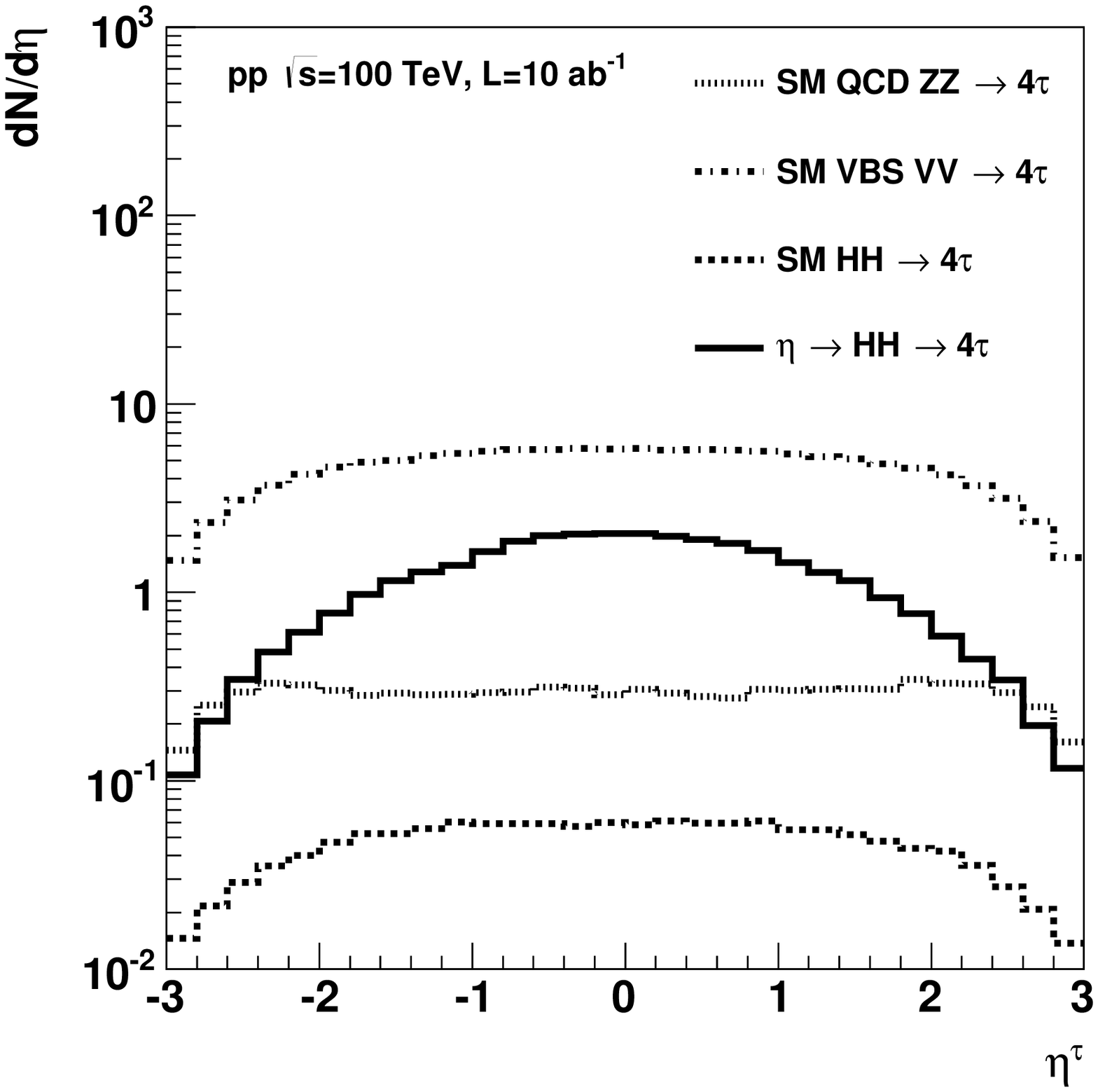}
  }
  \subfigure[The $p_T$ distribution of $\tau$-jets.]{
  \includegraphics[scale=0.45, angle=0]{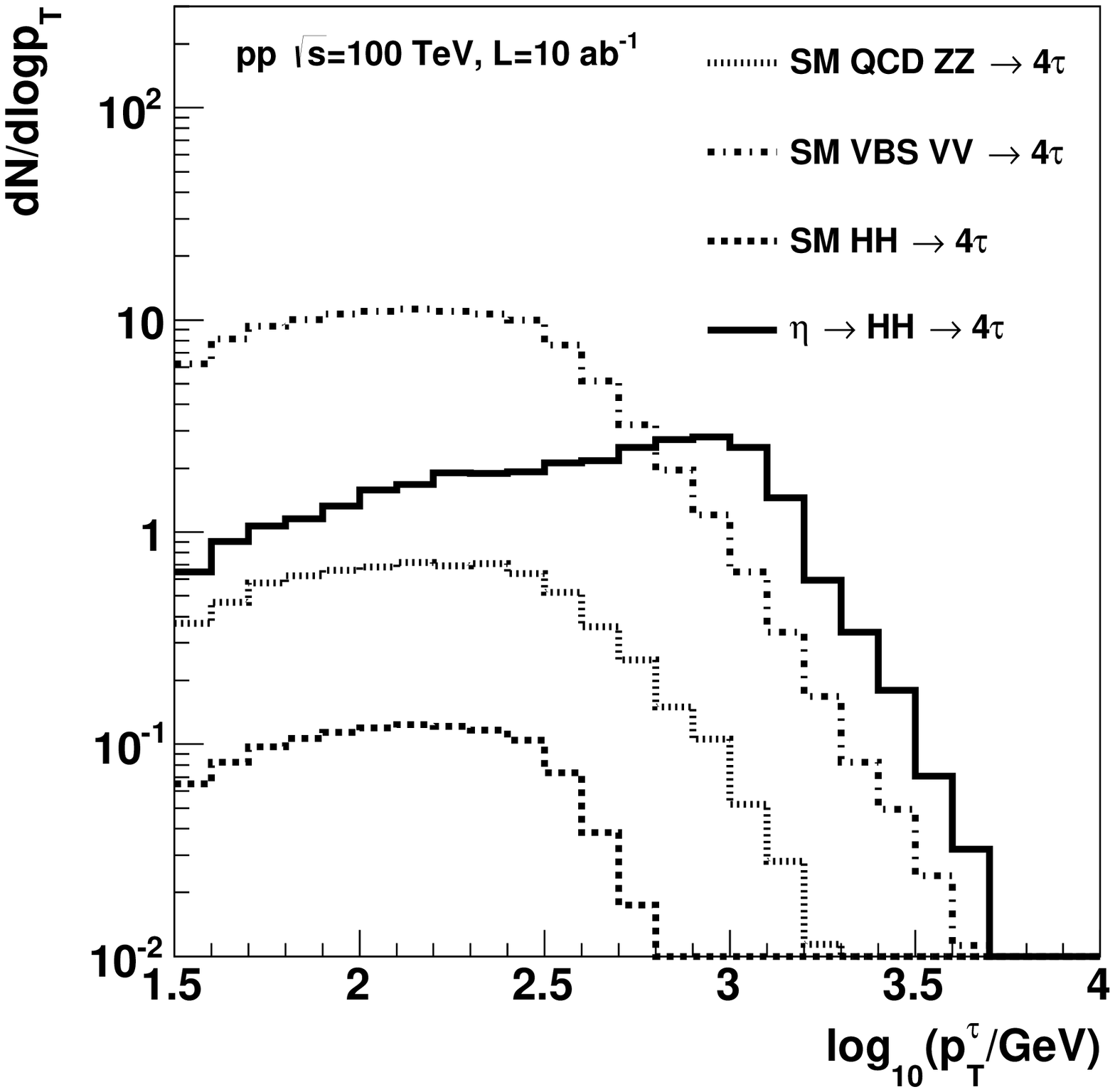}
  }
\caption{The pseudo-rapidity and $p_T$ distributions of $\tau$-jets. 
}
\label{tauPlots}
\end{figure}

    Figure~\ref{fj_eta} motivates the requirement $| \eta (\rm jet) | > 2$ to suppress diboson production via QCD processes without significant loss of signal efficiency, hence we apply this
  generator-level cut for the following distributions and studies. QCD showering by {\sc pythia} can generate additional jets in the central region.  The  distribution of the difference
  in pseudo-rapidities  $\Delta \eta_{jj} $ of the forward jets is shown in  Fig.~\ref{fj_deta}. This distribution motivates the additional selection requirement of $| \Delta \eta_{jj} | > 5$ 
 after {\sc pythia} showering. 

The scalar sum $H_T = \Sigma |\vec{p_T}| $ of all visible objects with pseudo-rapidity $|\eta| < 2.5$ 
in the event, called ``central $H_T$'' ($H_T^c$), is sensitive to QCD radiation accompanying the bosons and jets in the QCD-induced $ZZ + 2j$ process.  As shown in Fig.~\ref{fj_deta}, 
 the requirement $H_T^c < 300$~GeV suppresses
 this background. 

Figure~\ref{fj_pt} shows the distribution of the $p_T$ of the forward jets, and their $m_{jj}$ distribution. 
  The $p_T$ spectrum of the forward jets emitting longitudinal vector bosons due to the signal process $V_L V_L \to \eta$ is softer than the corresponding spectra from electroweak VBS and
 QCD processes, motivating the cut $p_T(\rm{jet}) < 500$~GeV to suppress the latter backgrounds. 
 
\begin{figure}[htp]
\centering
  \subfigure[The $p_T$ distribution of the leading $\tau$-jet.]{
  \includegraphics[scale=0.45, angle=0]{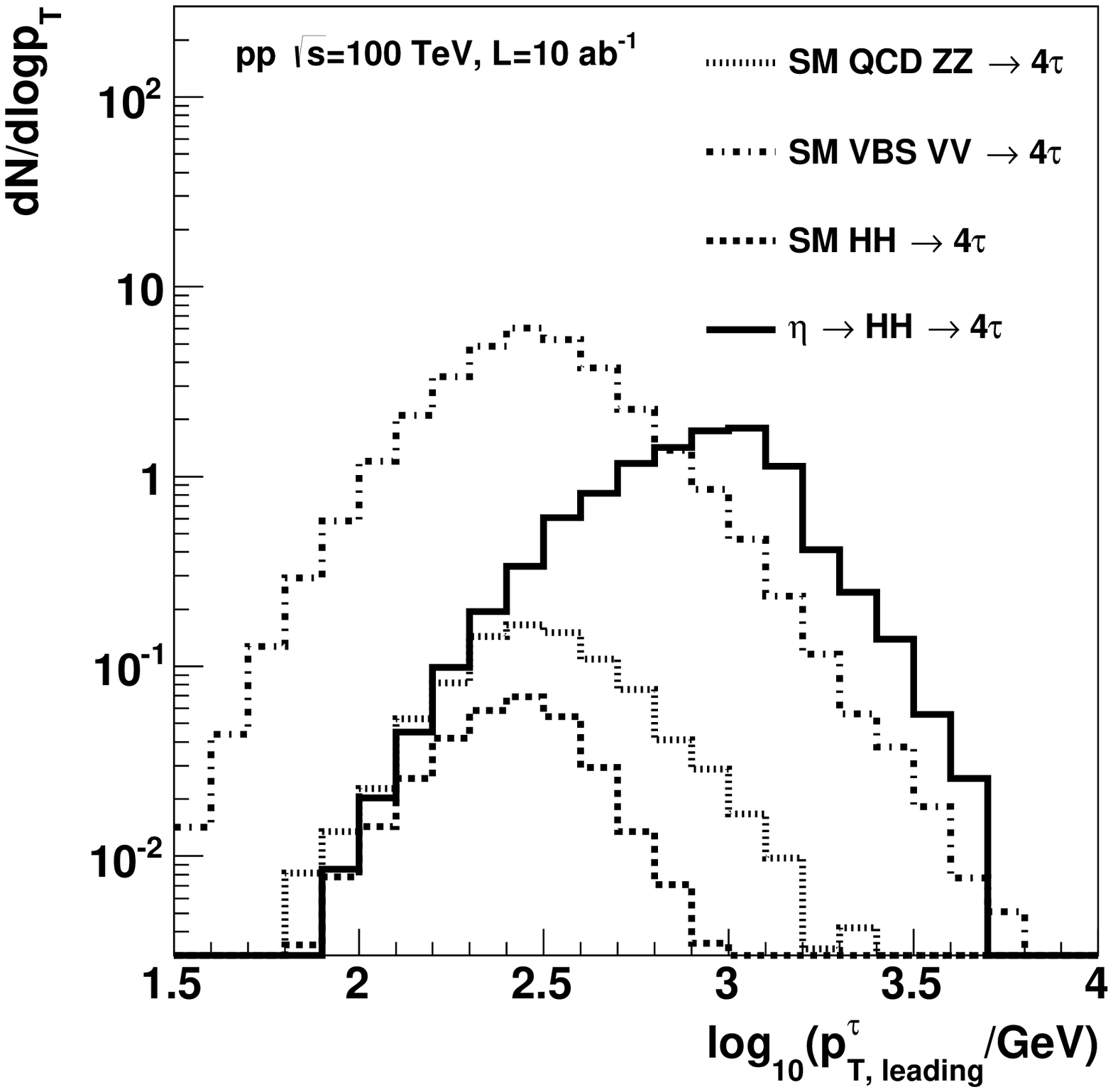}
  }
  \subfigure[The $p_T$ distribution of the next-to-leading $\tau$-jet. ]{
  \includegraphics[scale=0.45, angle=0]{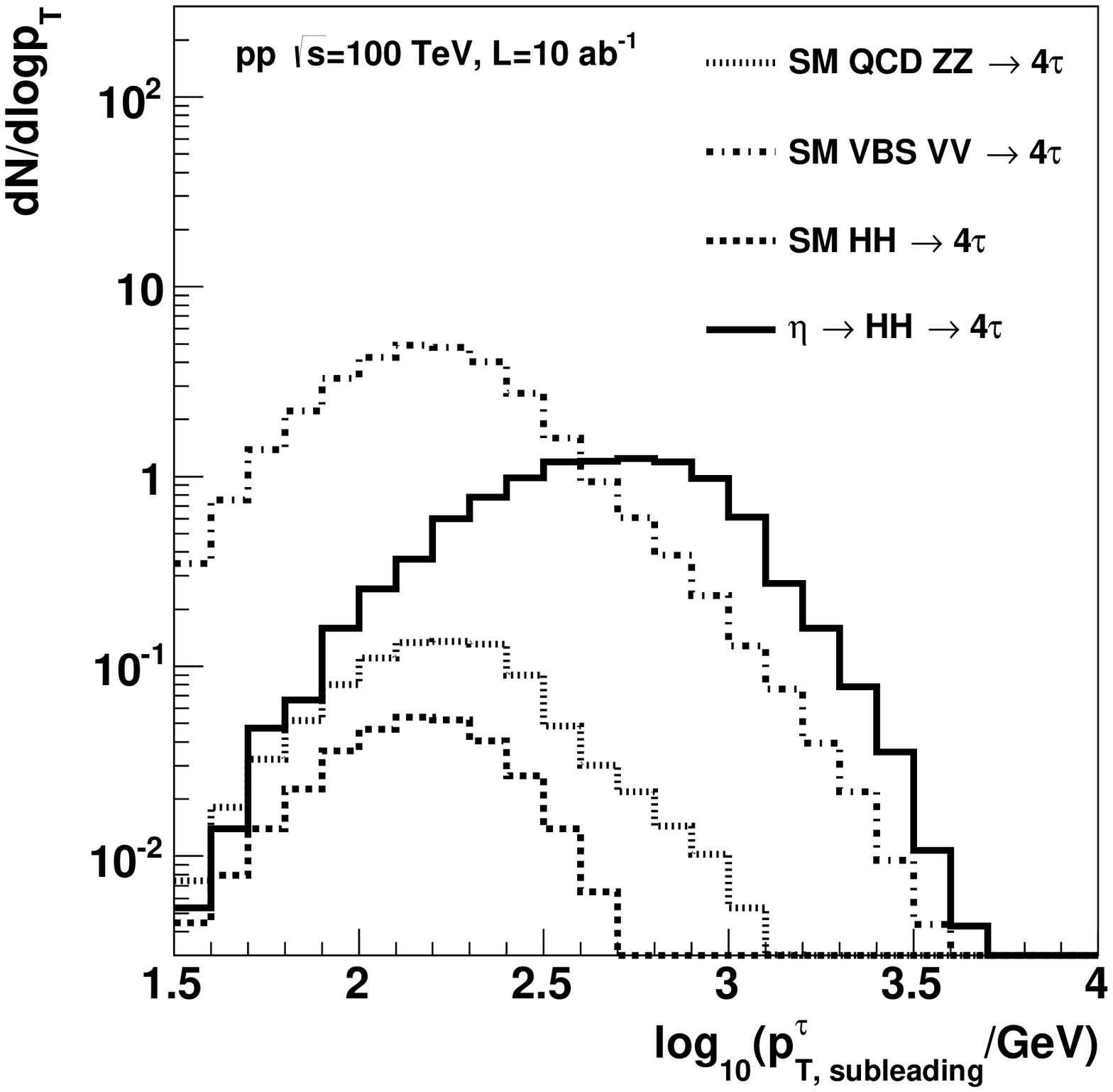}
  }
\caption{The $p_T$ distributions of the highest-$p_T$ $\tau$-jets.
}
\label{tauLeadingPlots}
\end{figure}

 The $\tau$ leptons from the $\eta \to HH \to 4\tau$ decay are produced more centrally and with higher $p_T$ than the backgrounds. We find that good coverage of $\tau$-jets up
  to $| \eta | < 3$ is adequate
 to have high  acceptance for the signal, as shown in Fig.~\ref{tauPlots}. The 
 inclusive $p_T$ spectrum of all $\tau$-jets is also shown in Fig.~\ref{tauPlots}. 
  The $p_T$ distributions of the leading and the next-to-leading $\tau$-jets, ranked in $p_T$, are shown in Fig.~\ref{tauLeadingPlots}. 

\begin{figure}[htp]
\centering
  \includegraphics[scale=0.45, angle=0]{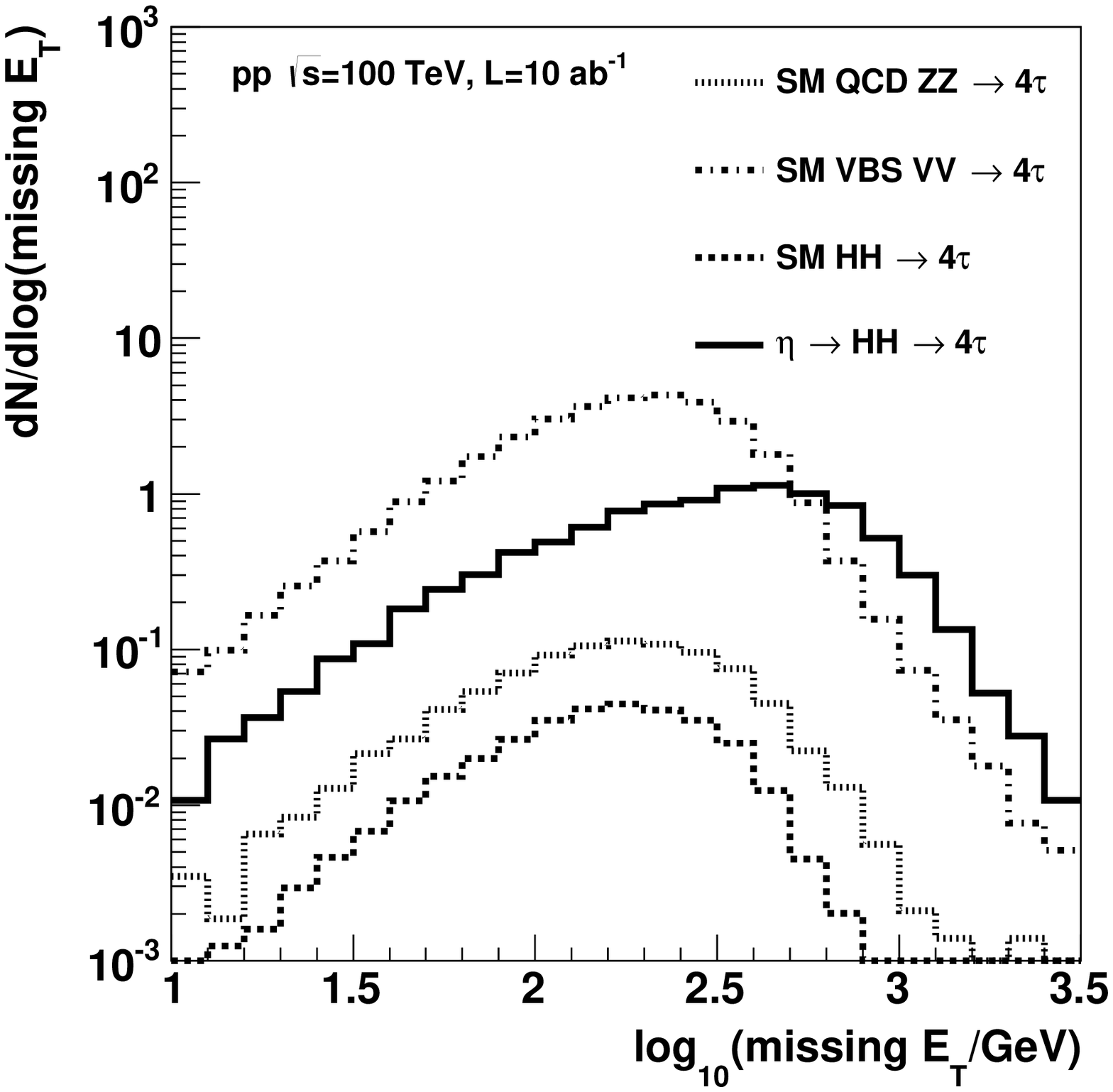}
\caption{The distributions of $\met$.
}
\label{metPlot}
\end{figure}

The vector sum $\met = | \Sigma \vec{p_T} |$ of all detected objects, which defines the missing transverse energy, 
  is sensitive to the angular correlations between the $\tau$-neutrinos emitted in the
 decay of the $\tau$ leptons. The $\tau$-leptons from Higgs boson decays have anti-parallel spins in the Higgs rest frame due to the Higgs being a scalar boson. Given the
 $V-A$ nature of the $\tau$-lepton decay vertex, the $\tau$ neutrinos are preferentially emitted parallel to each other in the rest frame, increasing the $\met$. In comparison, 
 the $\tau$ neutrinos are emitted anti-parallel to each other in $Z$ boson rest frame due to the unit spin of the latter, preferentially reducing the $\met$. The distribution of
 $\met$ is shown in Fig.~\ref{metPlot}. 

\begin{figure}[htp]
\centering
  \includegraphics[scale=0.45, angle=0]{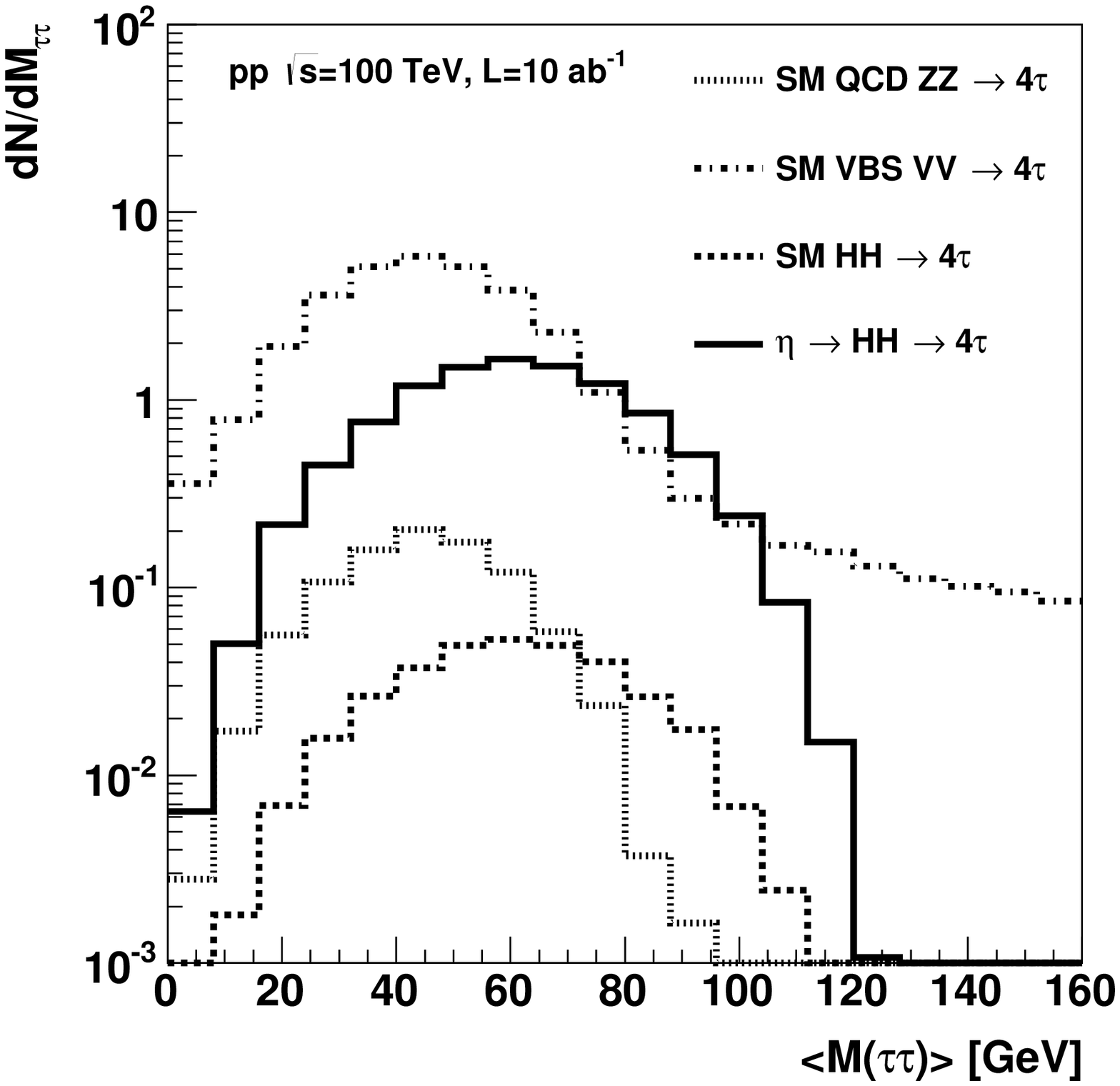}
\caption{The distribution of the per-event average of the reconstructed masses of the $\tau$-jet pairs from $V \to \tau \tau$ ($H \to \tau \tau$) in $VVjj$ ($HHjj$) events, 
 where $V = \gamma^* / Z$. }
\label{tauMassPlots}
\end{figure}

Additional discrimination between signal and background processes is provided by the invariant mass of combinations of $\tau$-jets. We  combine the $\tau$-jets
 from a given Higgs or $Z$ boson decay, and average the two resulting invariant masses in the event. The average reconstructed
 boson mass distribution is shown in Fig.~\ref{tauMassPlots}. The peak of the reconstructed Higgs boson mass is shifted to a higher value compared to the reconstructed $Z$ boson mass, as expected. 
 The use of sophisticated mass-reconstruction techniques that have been developed for di-$\tau$ resonances may be used to recover information lost with the neutrinos, which may compensate for 
 the experimental resolution on the visible momenta. Also shown in this figure are the distributions of the invariant mass all $\tau$-jets, and the combination of all $\tau$-jets and
 $\met$ setting the $\mathrm{E_z} \! \! \! \! \! \! / \; = 0$. 

 \begin{figure}[htp]
\centering
  \subfigure[The distribution of invariant mass of all reconstructed $\tau$-jets.]{
  \includegraphics[scale=0.45, angle=0]{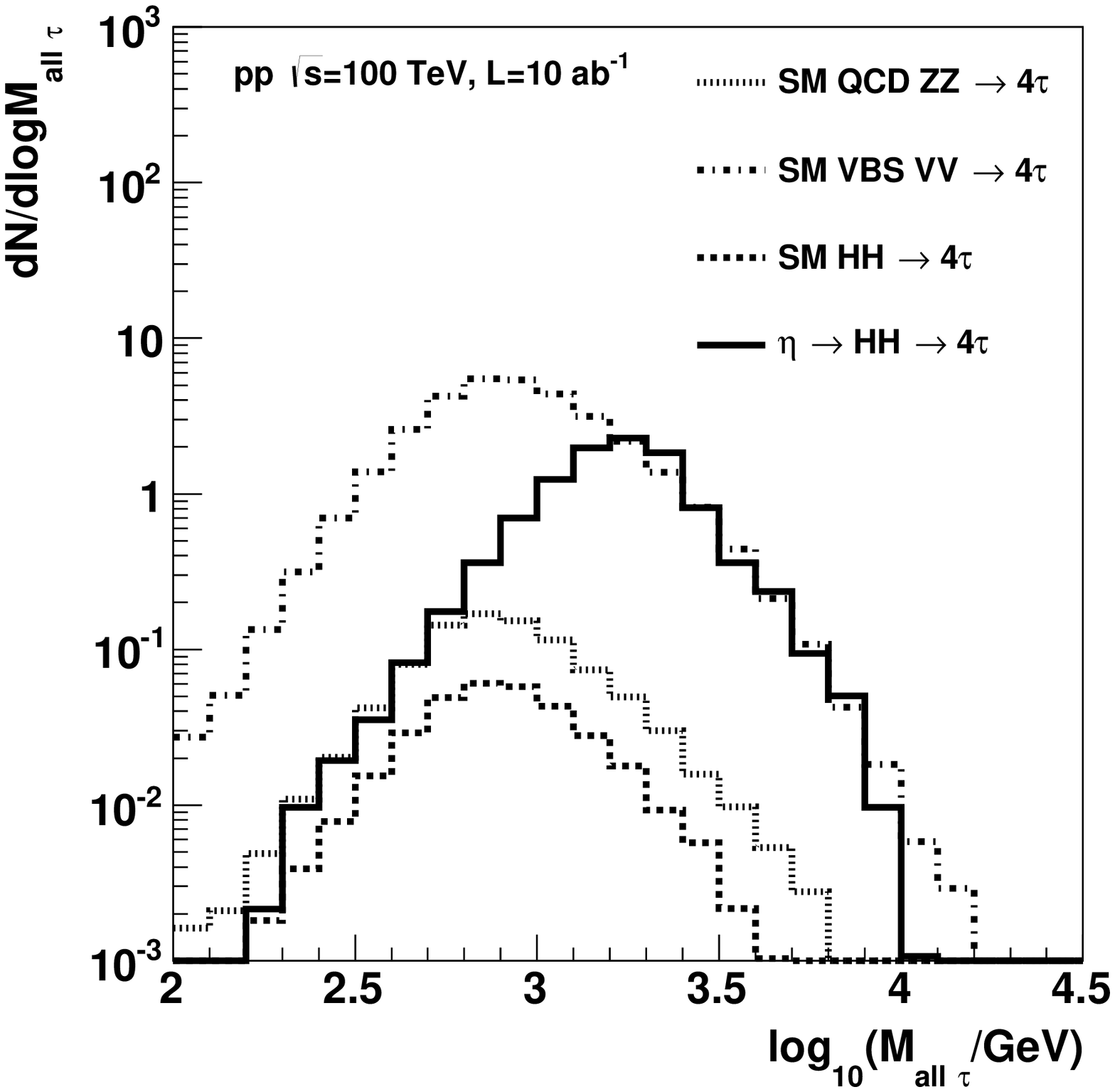}
  }
  \subfigure[The distribution of invariant mass of all reconstructed $\tau$-jets and $\met$, setting $E_z \! \! \! \! \! / \; = 0$. ]{
  \includegraphics[scale=0.45, angle=0]{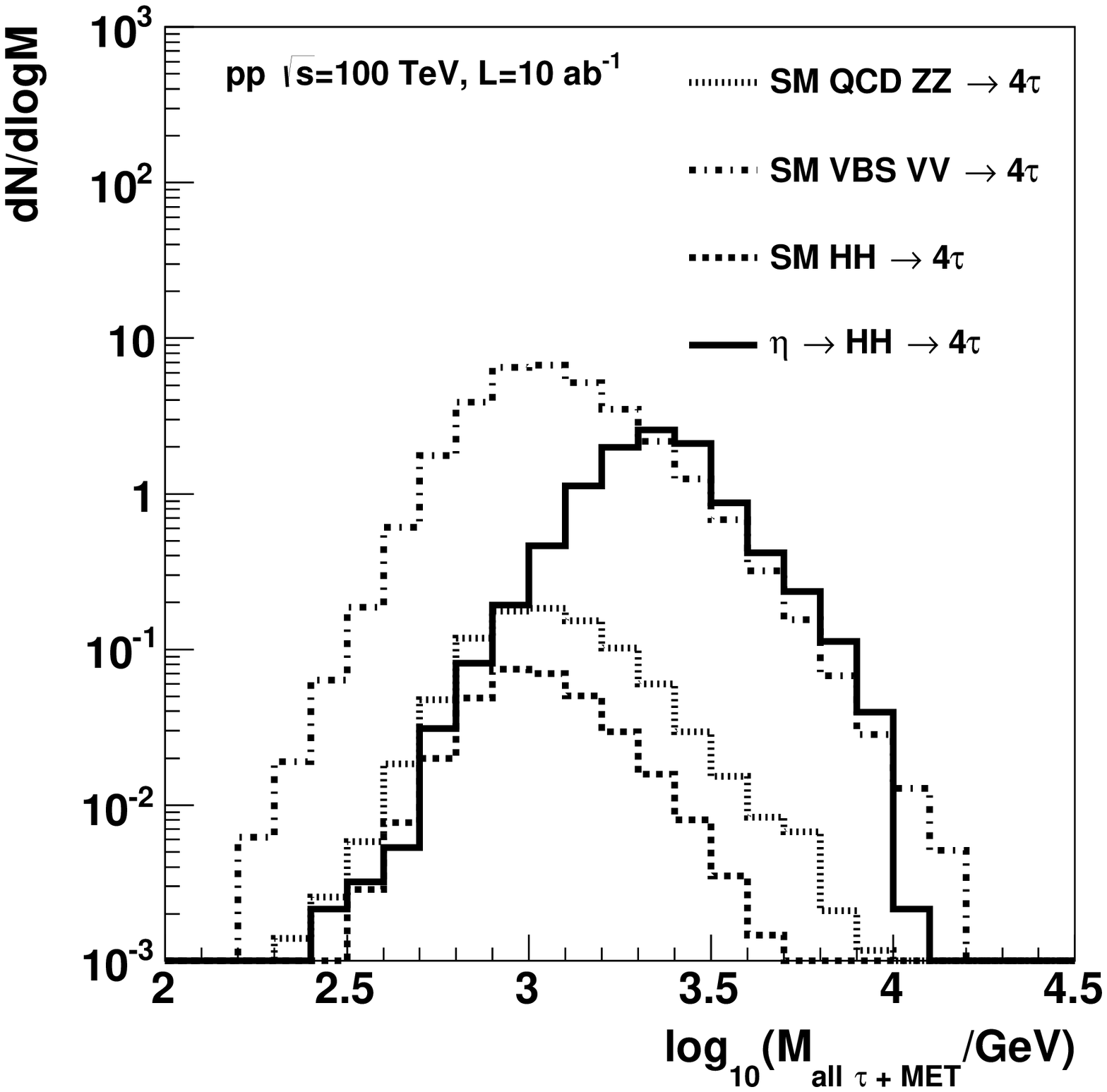}
  }
\caption{The invariant-mass distributions involving multiple reconstructed $\tau$-jets.
}
\label{tauMassPlots2}
\end{figure}

\section{Results}

\begin{figure}[htp]
\centering
\includegraphics[scale=0.47]{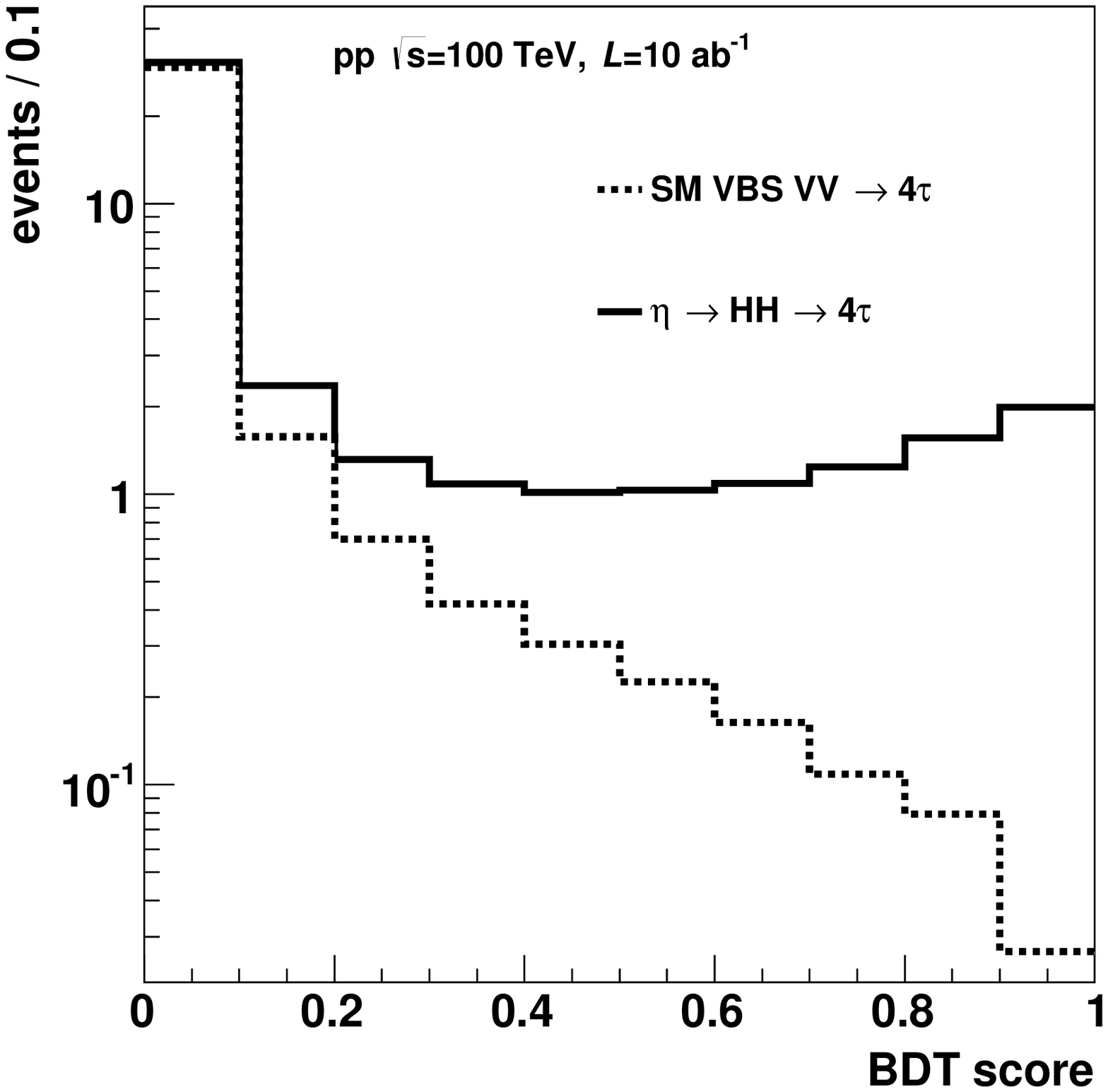}
\caption{The distributions of the BDT score for the $\eta \to HH \to 4 \tau$ signal with $m_\eta = 3$~TeV, and the VBS $VV \to 4 \tau$ background, where $V = \gamma^* / Z$. 
}
\label{nnHisto}
\end{figure}

After applying the selection cuts, the dominant irreducible background with the $4 \tau$ final state is $ZZ \to 4 \tau$ production in the VBS topology. We combine the information in the following
 distributions: the $p_T$ of the forward tagging jets and their pseudo-rapidity separation, the $p_T$ of the leading and sub-leading $\tau$ jets, the $\met$, the $H_T^c$, the average
 di-$\tau$ mass, the all-$\tau$ mass and the all-$\tau$+$\met$ mass, using a Boosted Decision Tree (BDT) algorithm to separate the $\eta \to HH \to 4 \tau$ signal from the VBS $ZZ \to 4 \tau$
 background. The resulting distributions of the BDT score for the signal and this dominant background are shown in Fig.~\ref{nnHisto}. We quantify the discovery reach for the signal 
 by computing the quantity $CL_b = P(Q<Q_{obs}|b)$, the probability for the test-statistic $Q$ to be smaller than the observed value given the background-only hypothesis. 
 When $1 - CL_b < 2.8 \times 10^{-7}$
 the background-only hypothesis is rejected at $5 \sigma$ significance. 
 The $5 \sigma$-discovery mass reach for the $\eta \to HH$ resonance for different fractional widths and integrated luminosities is shown
 in Table~\ref{etaReach}. 

\begin{table}[!ht]
\caption{$5 \sigma$ discovery mass reach for the $\eta \to HH \to 4 \tau$ resonance, at a $pp$ collider with $\sqrt{s} = 100$~TeV, as a function of integrated luminosity $\cal L$.}
\begin{ruledtabular}
\begin{tabular}{lccc}
 {$\cal L$} & \multicolumn{3}{c}{$m_\eta$ (TeV)} \\
(ab$^{-1}$) & $\Gamma/M = 5$\% & $\Gamma/M = 20$\% & $\Gamma/M = 70$\% \\
\hline
1  &  0.85\footnote{The minimum $p_T$ cuts on the $\tau$ lepton have been reduced for this mass point.} & 1.75  & 2.81 \\
3  &  1.33 & 2.25  & 3.42 \\
10 &  1.78 & 2.90  & 4.18 \\
30 &  2.30 & 3.56  & 4.94 \\
100 & 2.90 & 4.33  & 5.83 \\
\end{tabular}
\end{ruledtabular}
\label{etaReach}
\end{table}

 Table~\ref{ptReachCuts} shows the dependence of the $5\sigma$-discovery mass reach on the minimum $p_T$ cut applied on the forward tagging jets. The mass reach reduces by about 22\% for every
 20 GeV increase in the $p_T(\rm{jet})$ cut. Thus it is beneficial to maintain as low a $p_T(\rm {jet})$ cut as possible. Similarly, the forward rapidity coverage of jets is important. The
 dependence of the resonance mass reach as a function of the maximum jet rapidity detectable is shown in Table~\ref{etaReachCuts}. Coverage up to jet rapidity of 6-7 is desirable for a 100 TeV
 $pp$ collider. 
\begin{table}[!ht]
\caption{$5 \sigma$ discovery mass reach for the $\eta \to HH \to 4 \tau$ resonance, at a $pp$ collider with $\sqrt{s} = 100$~TeV and ${\cal L} = 10$~ab$^{-1}$, for various cuts values on minimum $p_T$ 
 of the forward jets. The fractional width of the $\eta$ resonance is set to $\Gamma/M = 20$\%.  }
\begin{ruledtabular}
\begin{tabular}{lccccc}
 $p_T^{\rm min}$ (GeV) & 30 & 50 & 70  & 90 & 110 \\
\hline
$m_\eta$ (TeV) & 3.53 &  2.90 &  2.35 & 1.92 & 1.56  
\end{tabular}
\end{ruledtabular}
\label{ptReachCuts}
\end{table}

\begin{table}[!ht]
\caption{$5 \sigma$ discovery mass reach for the $\eta \to HH \to 4 \tau$ resonance, at a $pp$ collider with $\sqrt{s} = 100$~TeV and ${\cal L} = 10$~ab$^{-1}$, for various cuts values on the maximum
 rapidity $(y)$  
 of the forward jets. The fractional width of the $\eta$ resonance is set to $\Gamma/M = 20$\%.  }
\begin{ruledtabular}
\begin{tabular}{lccccc}
 $y^{\rm max}$ & 8 & 7 & 6  & 5 & 4 \\
\hline
$m_\eta$ (TeV) & 2.9 & 2.9 & 2.81 & 2.42 & 1.75   
\end{tabular}
\end{ruledtabular}
\label{etaReachCuts}
\end{table}

\begin{table}[!ht]
\caption{$5 \sigma$ discovery mass reach for the $\eta \to HH \to 4 \tau$ resonance, as a function of the $\sqrt{s}$ of a $pp$ collider. The fractional resonance width $\Gamma_\eta / m_\eta$ 
is fixed at 20\%. These results are illustrated in Fig.~\ref{fiveSigmaTable20}.}
\begin{ruledtabular}
\begin{tabular}{lccc}
 {$\cal L$} & \multicolumn{3}{c}{$m_\eta$ (TeV)} \\
(ab$^{-1}$) & $\sqrt{s} = 50$~TeV & $\sqrt{s} = 100$~TeV & $\sqrt{s} = 200$~TeV \\
\hline
1   & 1.26 & 1.75  & 2.27 \\
3   & 1.58 & 2.25  & 2.88 \\
10  & 2.02 & 2.90  & 3.66 \\
30  & 2.49 & 3.56  & 4.44 \\
100 & 3.06 & 4.33  & 5.38 \\

\end{tabular}
\end{ruledtabular}
\label{energyReach}
\end{table}

\begin{figure}
\centering
\includegraphics[scale=0.57, angle=0]{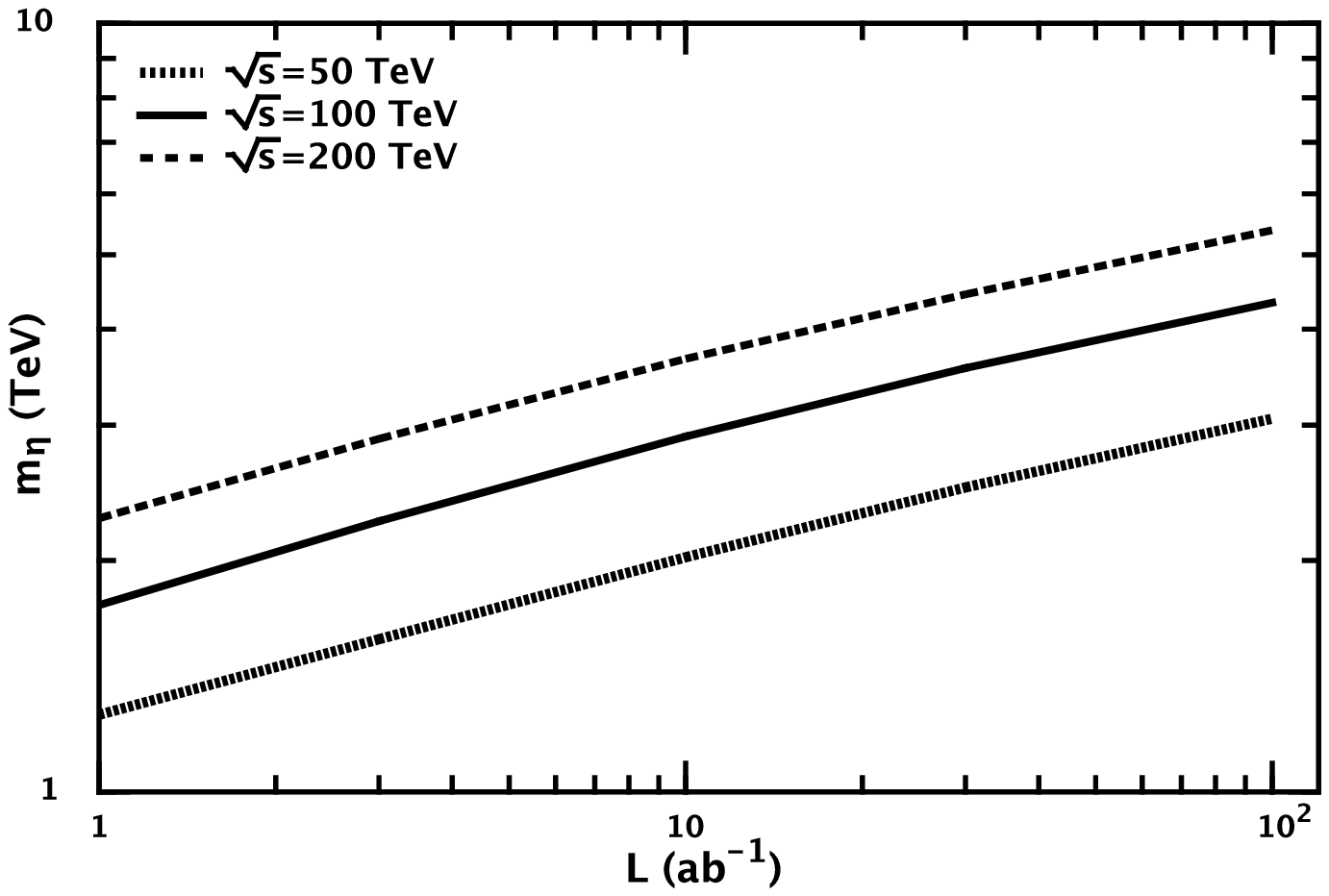}
\caption{$5 \sigma$ discovery mass reach for the $\eta \to HH \to 4 \tau$ resonance, as a function of the integrated luminosity and 
 $\sqrt{s}$ of a $pp$ collider. The fractional resonance width $\Gamma_\eta / m_\eta$
is fixed at 20\%. 
}
\label{fiveSigmaTable20}
\end{figure}

\begin{table}[!ht]
\caption{$5 \sigma$ discovery mass reach for the $\eta \to HH \to 4 \tau$ resonance, as a function of the $\sqrt{s}$ of a $pp$ collider. The fractional resonance width $\Gamma_\eta / m_\eta$ 
is fixed at 70\%. These results are illustrated in Fig.~\ref{fiveSigmaTable70}.}
\begin{ruledtabular}
\begin{tabular}{lccc}
 {$\cal L$} & \multicolumn{3}{c}{$m_\eta$ (TeV)} \\
(ab$^{-1}$) & $\sqrt{s} = 50$~TeV & $\sqrt{s} = 100$~TeV & $\sqrt{s} = 200$~TeV \\
\hline
1   & 1.89 & 2.81 & 3.85 \\
3   & 2.31 & 3.42 & 4.65 \\
10  & 2.83 & 4.18 & 5.63 \\
30  & 3.36 & 4.94 & 6.60 \\
100 & 3.97 & 5.83 & 7.74 \\

\end{tabular}
\end{ruledtabular}
\label{energyReachWider}
\end{table}

\begin{figure}
\centering
\includegraphics[scale=0.57, angle=0]{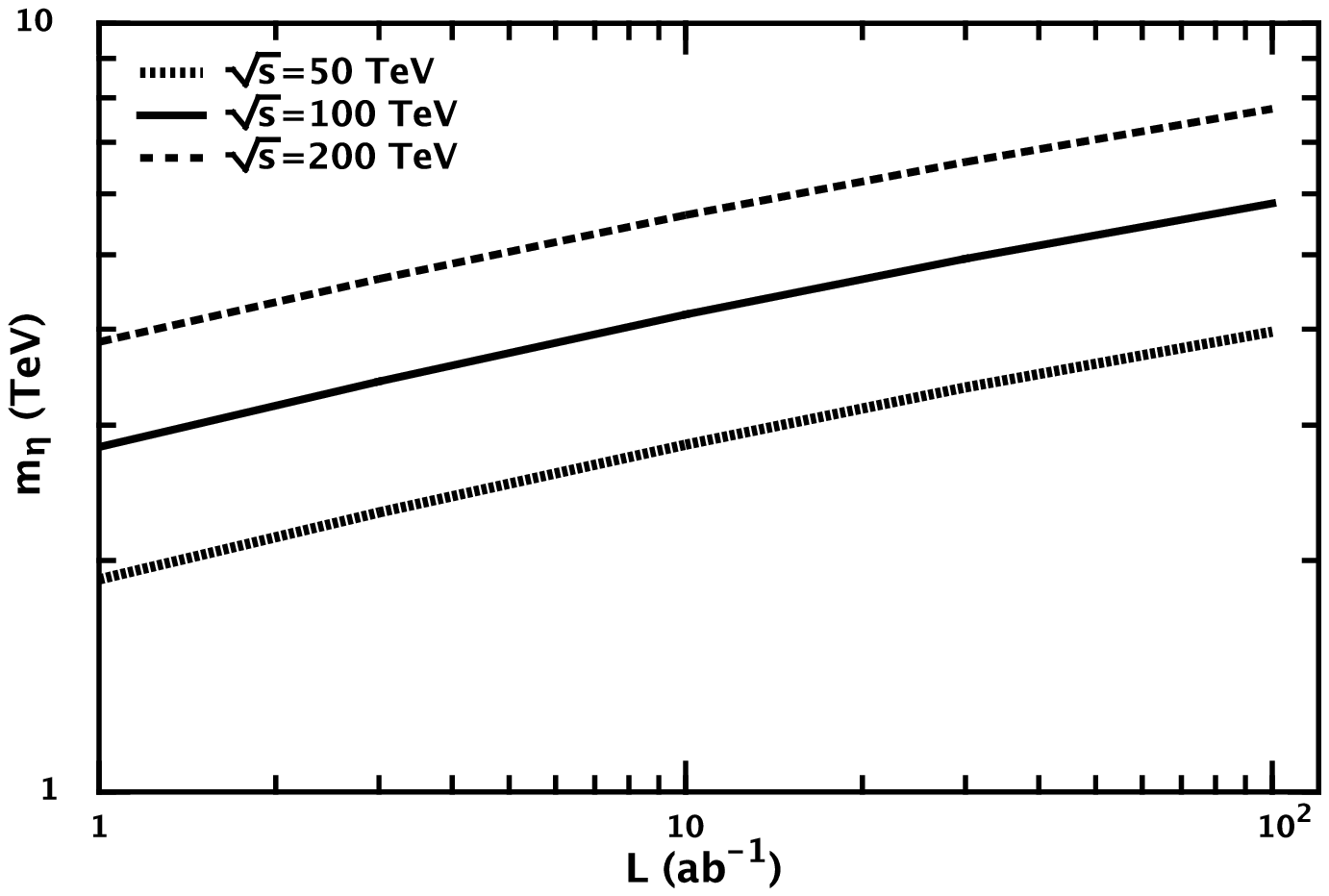}
\caption{$5 \sigma$ discovery mass reach for the $\eta \to HH \to 4 \tau$ resonance, as a function of the integrated luminosity and 
 $\sqrt{s}$ of a $pp$ collider. The fractional resonance width $\Gamma_\eta / m_\eta$
is fixed at 70\%.
}
\label{fiveSigmaTable70}
\end{figure}

Tables~\ref{energyReach} and~\ref{energyReachWider} summarize the discovery mass reach ($m_\eta^{5\sigma}$) as a function of integrated luminosity ($\cal L$)
  and collider center-of-mass energy $\sqrt {s}$, for resonance widths of 20\% and
 70\% respectively. The corresponding  results are also presented in Figs.~\ref{fiveSigmaTable20} and~\ref{fiveSigmaTable70}. The figures show that over a wide range of $\cal L$ and  $\sqrt {s}$,
  $m_\eta^{5\sigma}$ can be described fairly well by a power-law dependence on $\cal L$,
\begin{equation}
m_\eta^{5\sigma} \propto {\cal L}^\alpha
\end{equation}
where the $\alpha$ values are independent of $\sqrt {s}$. We find that the results of tables~\ref{energyReach} and~\ref{energyReachWider} can be parameterized by $\alpha = 0.20$~$(0.16)$ for a resonance width
 of 20\% (70\%). Equivalently, an increase in ${\cal L}$ by a factor of 10 raises the discoverable mass by 58\% (45\%).  
  The gain in mass reach with ${\cal L}$ is slightly more rapid at low ${\cal L}$ and slightly slower at high ${\cal L}$.  

We also attempt to describe the dependence of $m_\eta^{5\sigma}$ on $\sqrt{s}$ by a power law,
\begin{equation}
m_\eta^{5\sigma} \propto (\sqrt{s})^\beta
\end{equation}
and find that $\beta$ values are fairly independent of the integrated luminosity. If $\sqrt {s}$ is increased from 50 TeV to 100 TeV, $\beta = 0.50$~(0.56) for a fractional resonance width 
of 20\% (70\%). If  $\sqrt {s}$ is increased from 100 TeV to 200 TeV, the corresponding value of the power $\beta = 0.34$~(0.43) fits the results. Thus, the scaling behavior 
 for the discovery mass reach as a function of $\sqrt{s}$ can be approximated by a power-law behavior but the sensitivity gain starts to saturate at the higher collider 
 energies. A reasonable approximation 
 is  obtained by fitting the gain from 50 TeV to 200 TeV, yielding $\beta = 0.42$~(0.50) for $\Gamma / M = $20\% (70\%). Equivalently, a doubling of the collider energy increases the discovery
 mass reach by 33-40\% depending on the resonance width, with a somewhat larger (smaller) increase at lower (higher) energies.  

 It is interesting to evaluate the trade-off between collider energy and integrated luminosity for a given discovery mass reach. For a fractional resonance width of 20\% (70\%), a 
factor of two in collider energy is equivalent to a factor of 4.3 (8.7) in integrated luminosity. For a narrow (implying weakly coupled)
  resonance, integrated luminosity is more effective while for a wide (implying strongly coupled) resonance, collider energy is more effective as a means of increasing the mass reach. 
 
 Studies of the sensitivity of the HL-LHC for resonances in vector 
         boson scattering indicate a discovery potential of $m_{\rm res}/g_{\rm res}$  
 of $\approx 500$~GeV~\cite{atlasHLLHCvbs}, where $g_{\rm res}$ is the resonance coupling. While the resonance model and decay channel used in that study were different from 
 ours, the  sensitivity from Table~\ref{etaReach} is about a factor of 4-7 higher than that at the LHC.  Comparative studies between the HL-LHC and a 100
           TeV $pp$ collider~\cite{snowmassEWKreport} regarding the sensitivity to dimension-8 operators 
           in vector boson scattering can also be interpreted as a factor of four higher
           mass scale being probed at the 100 TeV collider.

\section{Conclusions}
 The sensitivity to a resonance in longitudinal VBS in the mass range of 1.5-5~TeV 
 decaying to $HH \to 4 \tau$ is discussed for a 100 TeV  $pp$ collider.  In a benchmark model motivated by the spontaneous breaking of a global $SO(5)$ symmetry to $SO(4)$, the $SU(2)_L$ Higgs
 doublet field contains the set of four Goldstone modes, which are derivatively coupled to this resonance. The resonance decays democratically to longitudinal $W^\pm$ and $Z$ bosons and Higgs bosons 
 with 2:1:1 proportion for the respective branching ratios. 
 We have used leading-order cross sections for the signal and background processes, as the dominant background of $VV$ production ($V = Z, \gamma^*$) in the VBS topology is also
 purely electroweak and higher-order QCD corrections can be expected to increase the signal and background cross section in comparable  proportion. 

 A reconstruction efficiency for $\tau$ leptons of 60\% and a corresponding QCD jet efficiency of a few percent
 is assumed in this study. Future detectors are expected to maintain this $\tau$ reconstruction performance, currently achieved by the LHC experiments, at higher transverse momenta with 
 acceptance up to a pseudo-rapidity of 3. 

 This study shows that, for an integrated luminosity of 10 ab$^{-1}$ at a $pp$ collider with $\sqrt{s} = 100$~TeV, 
 a $5 \sigma$ discovery reach of 2.90 TeV for the mass of the resonance can be achieved, assuming a width of 20\%. For widths varying between
 5\% and 70\%, the corresponding mass reach varies from 1.78 TeV and 4.18 TeV. For a factor of three increase in integrated luminosity, the mass reach increases by 25\% (19\%) for 
 fractional resonance width of 20\% (70\%), and this power-law scaling behavior is independent of the collider energy. 
 An approximate power-law scaling dependence on $\sqrt{s}$ is also found, where a doubling of the collider energy increases the discovery
 mass reach by 33-40\% depending on the resonance width, with a larger proportionate increase at lower energies.
 
The trade-off between collider energy and integrated luminosity for a given discovery mass reach favors luminosity for a narrow, weakly-coupled resonance and energy for a wide, strongly-coupled resonance.
 For a fractional resonance width of 20\% (70\%), a 
factor of two in collider energy is equivalent to a factor of 4.3 (8.7) in integrated luminosity. 
 
 New physics in longitudinal vector boson scattering puts strong requirements on the detection of forward jets at high rapidities and low $p_T$. The mass reach is reduced by about 22\% for every
 20 GeV increase in the minimum $p_T(\rm{jet})$ requirement on the forward tagging jets. A minimum rapidity coverage for the forward jets of 6-7 is desirable. For $p_T ({\rm jet}) > 50$~GeV, the 
 mass reach is reduced by 14\% if the rapidity coverage is reduced from 6 to 5. Further reducing the rapidity coverage to 4 causes the mass  reach to drop by another 28\% relative to the coverage
 of 5. If VBS jets could be distinguished from pileup jets at lower $p_T$, the gains from extended rapidity coverage would be even higher. 

This paper also highlights the importance of the $H \to \tau \tau$ decay channel as a relatively clean mode of identification for double-Higgs production. Distinguishing highly-boosted $\tau$ leptons
 with $p_T(\tau) \sim 1$~TeV from QCD jets and electrons presents a challenge that could be addressed by high-granularity electromagnetic calorimeters. Used in conjunction with tracking detectors 
  having good two-track resolution, 
such calorimeters could  measure the individual charged particles and photons within the $\tau$-jets with sufficient spatial and energy resolution. Resolving the substructure in $\tau$-jets 
 and possibly extracting information on $\tau$ polarization could allow future detectors to maintain or even surpass the $\tau$ identification efficiency 
 and background rejection that has  been demonstrated by the LHC experiments at lower transverse momenta. Thus, the $H \to \tau \tau$ mode could provide a discovery channel for resonant double-Higgs
 production with good signal-to-background ratio, albeit with low statistics, similar to the $H \to ZZ \to 4 \ell$ ($\ell = e, \mu$) channel for the Higgs boson discovery at the LHC. The
 fact that the $H \to \tau \tau$  branching ratio is approximately equal to the combined branching rato for $Z \to ee, \mu \mu$ supports our emphasis on $\tau$ detection. 
\section*{Acknowledgments}
We thank  Kaustubh Agashe, Nima Arkani-Hamed, 
 Roberto Contino, Estia Eichten,  
Elisabetta Furlan, Zhen Liu, Michelangelo Mangano, Giuliano Panico,  Chris Quigg, Raman Sundrum,  
 Liantao Wang, Andrea Wulzer,  and  Felix Yu  
  for helpful discussions. We thank Benjamin Cerio for his help with the BDT software.  
 The research of S.~C. was supported by the Department of Energy Contract No. DE-AC02-06CH11357 at Argonne National Laboratory. The work of A.~K. was supported by the Fermi National Accelerator
 Laboratory and by a Department of Energy grant to Duke University. 
 Fermilab is operated by Fermi Research Alliance, LLC, under Contract No. DE-AC02-07CH11359 with the United States Department of Energy.

\clearpage
\bibliography{biblio}

\appendix

\begin{widetext}
\onecolumngrid

\section*{Appendix}
\vspace*{1mm} 

\begin{table}[h]
\begin{tabular}{l|cc}
Decay channel & Branching ratio & Uncertainty \\ \hline \\
$b \bar{b} b \bar{b}$                 &   $3.33\cdot 10^{-1}$   &   $\pm\> 2.20\cdot 10^{-2}$   \\
$\tau \tau b \bar{b}$                 &   $7.29\cdot 10^{-2}$   &   $\pm\> 4.80\cdot 10^{-3}$   \\
$W^{+} (\to l \nu)  W^{-} (\to l \nu) b \bar{b}$   &   $1.09\cdot 10^{-2}$   &   $\pm\> 5.93\cdot 10^{-4}$   \\
$\tau \tau \tau \tau$                 &   $3.99\cdot 10^{-3}$   &   $\pm\> 4.55\cdot 10^{-4}$   \\
$\gamma \gamma b \bar{b}$             &   $2.63\cdot 10^{-3}$   &   $\pm\> 1.58\cdot 10^{-4}$   \\
$W^{+} (\to l \nu)  W^{-} (\to l \nu) \tau \tau$   &   $1.20\cdot 10^{-3}$   &   $\pm\> 8.56\cdot 10^{-5}$   \\
$\gamma \gamma \tau \tau$             &   $2.88\cdot 10^{-4}$   &   $\pm\> 2.19\cdot 10^{-5}$   \\
$b \bar{b} \mu^{+}\mu^{-}$            &   $2.53\cdot 10^{-4}$   &   $\pm\> 1.73\cdot 10^{-5}$   \\
$Z (\to l^{+} l^{-} ) Z(\to l^{+} l^{-}) b \bar{b}$   &   $1.41\cdot 10^{-4}$   &   $\pm\> 7.64\cdot 10^{-6}$   \\
$b \bar{b} Z (\to l^{+} l^{-} ) \gamma$   &   $1.21\cdot 10^{-4}$   &   $\pm\> 1.16\cdot 10^{-5}$   \\
$W^{+} (\to l \nu)  W^{-} (\to l \nu) W^{+} (\to l \nu)  W^{-} (\to l \nu)$   &   $8.99\cdot 10^{-5}$   &   $\pm\> 7.73\cdot 10^{-6}$   \\
$\gamma \gamma W^{+} (\to l \nu)  W^{-} (\to l \nu)$   &   $4.32\cdot 10^{-5}$   &   $\pm\> 2.85\cdot 10^{-6}$   \\
$\tau \tau \mu^{+}\mu^{-}$            &   $2.77\cdot 10^{-5}$   &   $\pm\> 2.29\cdot 10^{-6}$   \\
$Z (\to l^{+} l^{-} ) Z(\to l^{+} l^{-}) \tau \tau$   &   $1.54\cdot 10^{-5}$   &   $\pm\> 1.10\cdot 10^{-6}$   \\
$\tau \tau Z (\to l^{+} l^{-} ) \gamma$   &   $1.32\cdot 10^{-5}$   &   $\pm\> 1.41\cdot 10^{-6}$   \\
$\gamma \gamma \gamma \gamma$         &   $5.20\cdot 10^{-6}$   &   $\pm\> 5.20\cdot 10^{-7}$   \\
$W^{+} (\to l \nu)  W^{-} (\to l \nu) \mu^{+}\mu^{-}$   &   $4.15\cdot 10^{-6}$   &   $\pm\> 3.07\cdot 10^{-7}$   \\
$Z (\to l^{+} l^{-} ) Z(\to l^{+} l^{-}) W^{+} (\to l \nu)  W^{-} (\to l \nu)$   &   $2.31\cdot 10^{-6}$   &   $\pm\> 1.41\cdot 10^{-7}$   \\
$W^{+} (\to l \nu)  W^{-} (\to l \nu) Z (\to l^{+} l^{-} ) \gamma$   &   $1.99\cdot 10^{-6}$   &   $\pm\> 1.98\cdot 10^{-7}$   \\
$\gamma \gamma \mu^{+}\mu^{-}$        &   $9.99\cdot 10^{-7}$   &   $\pm\> 7.80\cdot 10^{-8}$   \\
$\gamma \gamma Z (\to l^{+} l^{-} ) Z(\to l^{+} l^{-})$   &   $5.57\cdot 10^{-7}$   &   $\pm\> 3.67\cdot 10^{-8}$   \\
$\gamma \gamma Z (\to l^{+} l^{-} ) \gamma$   &   $4.78\cdot 10^{-7}$   &   $\pm\> 4.92\cdot 10^{-8}$   \\
$Z (\to l^{+} l^{-} ) Z(\to l^{+} l^{-}) \mu^{+}\mu^{-}$   &   $5.35\cdot 10^{-8}$   &   $\pm\> 3.95\cdot 10^{-9}$   \\
$Z (\to l^{+} l^{-} ) \gamma \mu^{+}\mu^{-}$   &   $4.59\cdot 10^{-8}$   &   $\pm\> 4.96\cdot 10^{-9}$   \\
$Z (\to l^{+} l^{-} ) Z(\to l^{+} l^{-}) Z (\to l^{+} l^{-} ) \gamma$   &   $2.56\cdot 10^{-8}$   &   $\pm\> 2.55\cdot 10^{-9}$   \\
$Z (\to l^{+} l^{-} ) Z(\to l^{+} l^{-}) Z (\to l^{+} l^{-} ) Z(\to l^{+} l^{-})$   &   $1.49\cdot 10^{-8}$   &   $\pm\> 1.28\cdot 10^{-9}$   \\
$Z (\to l^{+} l^{-} ) \gamma Z (\to l^{+} l^{-} ) \gamma$   &   $1.10\cdot 10^{-8}$   &   $\pm\> 1.97\cdot 10^{-9}$   \\
\end{tabular}
\caption{Branching ratios for final states arising from double-Higgs production, with the requirement of leptonic decays of $W$ and $Z$ bosons. }
\label{branchingRatios}
\end{table}

\begin{figure}
\centering
\includegraphics[scale=0.35]{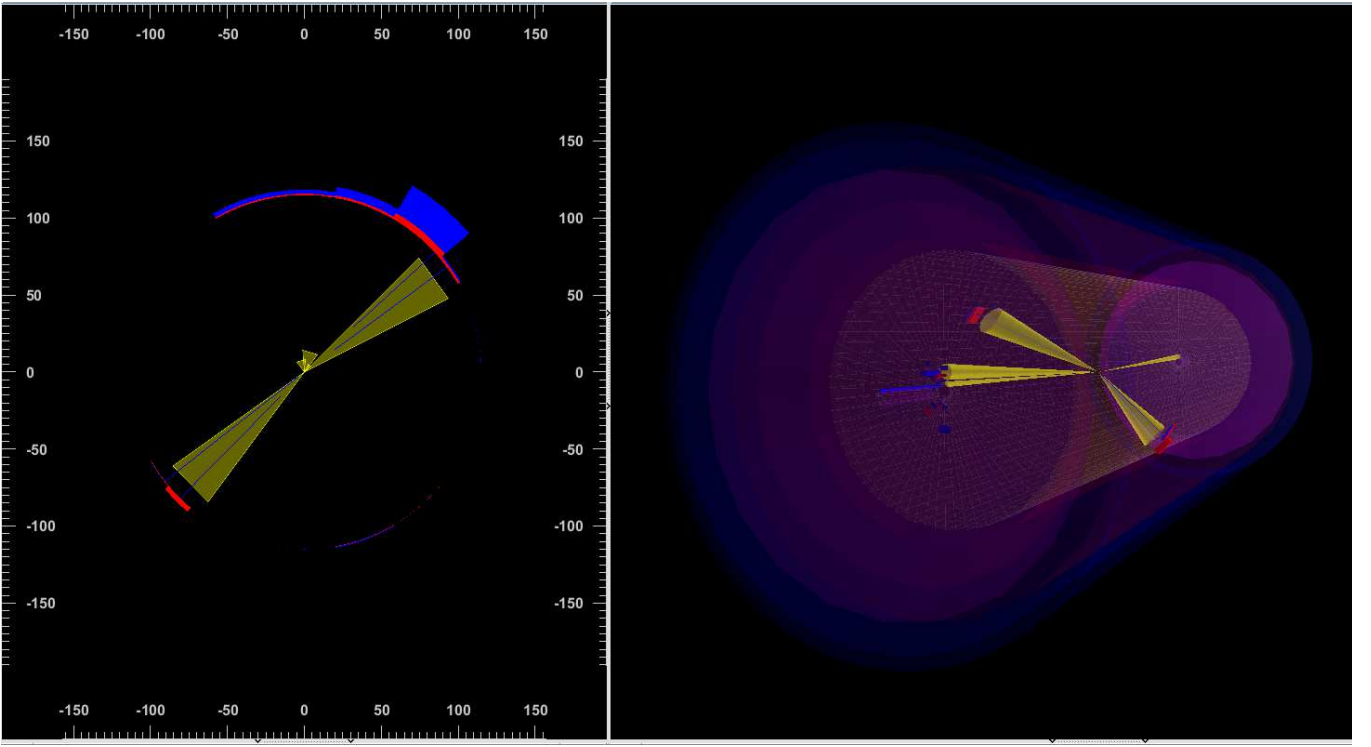}
\caption{An event display showing the $\eta$ resonance ($m_\eta  = 3$~TeV) produced via longitudinal vector boson scattering, and decaying to $HH \to 4\tau$ at
a $100$~TeV $pp$ collider.
(left) In the transverse view, two 
 forward jets with transverse momenta above 50~GeV are shown with the small yellow cones, and two jets arising from $H \to \tau \tau$ are shown with the large yellow cones. (right) A 3D view
 of the same event. The jets are
reconstructed using the anti-$k_T$ algorithm \cite{Cacciari:2008gp}
  using the {\sc FastJet} package~\cite{fastjet}.
The event display was created using the {\sc Delphes}
fast simulation \cite{deFavereau:2013fsa},
{\sc HepSim} \cite{Chekanov:2014fga} and the Snowmass detector setup \cite{Anderson:2013kxz}.
The blue lines show charged hadrons.
See the text for details.
}
\label{View2}
\end{figure}
\end{widetext}

\end{document}